\documentclass[journal,10pt]{IEEEtran}

\usepackage{amsmath,amssymb,amscd}
\usepackage{graphicx,cite}
\usepackage{algorithm}
\usepackage{algpseudocode}
\usepackage{subfigure}
\usepackage{multirow}
\usepackage[english]{babel}
\usepackage[autostyle]{csquotes}
\usepackage[version=3]{mhchem}
    
\newcommand{\m}{\text{m}}
\newcommand{\cc}{\text{c}}

\newcommand{\oo}{\text{O}}
\newcommand{\nn}{\text{N}}

\newtheorem{lemma}{Lemma}

\begin{document}
\title{Mobile Edge Computing via a UAV-Mounted Cloudlet: Optimization of Bit Allocation and Path Planning}
\author{\IEEEauthorblockN{Seongah Jeong, Osvaldo Simeone, and Joonhyuk Kang}\\

\thanks{
S. Jeong is with the School of Engineering and Applied Sciences (SEAS), Harvard University, 29 Oxford street, Cambridge,
MA 02138, USA (e-mail: sej293@g.harvard.edu).

O. Simeone is with the Department of Informatics, King's College London, Strand, London WC2R 2LS, England, UK (e-mail: osvaldo.simeone@kcl.ac.uk).

J. Kang is with the Department of Electrical Engineering, Korea Advanced Institute of Science and Technology
(KAIST), 291, Daehak-ro, Yuseong-gu, Daejeon 34141, Korea (e-mail: jhkang@ee.kaist.ac.kr).}
}
\maketitle
\begin{abstract}
Unmanned Aerial Vehicles (UAVs) have been recently considered as means to provide enhanced coverage or relaying services to mobile users (MUs) in wireless systems with limited or no infrastructure. In this paper, a UAV-based mobile cloud computing system is studied in which a moving UAV is endowed with computing capabilities to offer computation offloading opportunities to MUs with limited local processing capabilities. The system aims at minimizing the total mobile energy consumption while satisfying quality of service requirements of the offloaded mobile application. Offloading is enabled by uplink and downlink communications between the mobile devices and the UAV that take place by means of frequency division duplex (FDD) via orthogonal or non-orthogonal multiple access (NOMA) schemes. The problem of jointly optimizing the bit allocation for uplink and downlink communication as well as for computing at the UAV, along with the cloudlet's trajectory under latency and UAV's energy budget constraints is formulated and addressed by leveraging successive convex approximation (SCA) strategies. Numerical results demonstrate the significant energy savings that can be accrued by means of the proposed joint optimization of bit allocation and cloudlet's trajectory as compared to local mobile execution as well as to partial optimization approaches that design only the bit allocation or the cloudlet's trajectory.  
\end{abstract}

\section{Introduction}\label{sec:intro}
The deployment of moving base stations or relays mounted on unmanned aerial vehicles (UAVs) is a promising solution to extend the coverage of a wireless system to areas in which there is a limited available infrastructure of wireless access points, such as in developing countries or rural environments, as well as in disaster response, emergency relief and military scenarios \cite{Facebook, Google, JSA17IET,  Loke15Arxiv, Zeng16Comm, Asadpour13ACM, Asadpour14Commag}. However, the limited coverage and mobility of energy-constrained UAVs introduce new challenges for the design of UAV-based wireless communications. As a result, recent research activity has focused on the problems of path planning and energy-aware deployment for UAV-based systems \cite{Zeng16TCOM, Zhao04ACM, Shah03Adhoc, Zhan11TAES, Soorki16Arxiv, Mozaffari15Arxiv, Mozaffari16Arxiv, Mozaffari16Arxiv2, Zeng16Arxiv2, Manyam16Arxiv, Dorling16TSMC}, as we briefly review below.

\subsection{State of the Art}\label{sec:art}
In \cite{Zeng16TCOM, Zhao04ACM, Shah03Adhoc, Zhan11TAES}, a UAV-enabled mobile relaying system is studied, where the role of the UAV is to act as a \textit{relay} for communication between wireless devices. In particular, the problem of jointly optimizing the power allocation at source and moving relay, as well as the relay's trajectory, is tackled in \cite{Zeng16TCOM} with the aim of maximizing the throughput under mobility constraints on the relay's speed and terminal locations and assuming a decode-store-and-forward scheme. To address the problem, an iterative algorithm is proposed to alternatively optimize the power allocation and relay's trajectory. In \cite{Zhao04ACM}, the problem of efficient data delivery in sparse mobile \textit{ad hoc} networks is studied, where a set of moving relays between pairs of sources and destinations is employed. Two types of relaying schemes are developed in order to minimize the message drop rate under energy constraints, whereby either the nodes move to meet a given relay's trajectory, or a relay moves to meet static nodes. Both schemes are optimized in terms of trajectory of either the nodes or the relay. A similar scheme has also been introduced for sparse \textit{sensor} networks in \cite{Shah03Adhoc}.

The authors in \cite{Zhan11TAES} study the deployment of UAVs acting as relays between ground terminals and a network base station so as to provide \textit{uplink} transmission coverage for ground-to-UAV communication. The problem of optimizing the UAV heading angle is tackled with the goal of maximizing the sum-rate under individual minimal rate constraints. To this end, the authors derive a closed-form expression  approximate for the average uplink data rate for each link. In \cite{Soorki16Arxiv}, a scheduling and resource allocation framework is developed for energy-efficient \textit{machine-to-machine} communications with UAVs, where multiple UAVs provide uplink transmission to collect the data from the heads of the clusters consisting of a number of machine-type devices. The authors investigate the minimum number of required UAVs to serve the cluster heads and their dwelling time over each cluster head by using the queue rate stability concept. 

References \cite{Mozaffari15Arxiv, Mozaffari16Arxiv}, instead, study the optimal deployment of multiple UAVs acting as \textit{flying base station}s in the \textit{downlink} scenario. The optimal altitude for a single UAV is addressed with the aim of minimizing the required downlink transmit power for covering a target area, and then the treatment is extended to two UAVs with and without  interference between the UAVs in \cite{Mozaffari15Arxiv}. In contrast, in \cite{Mozaffari16Arxiv}, the minimization of the total required downlink transmit power from the UAVs is tackled under minimum users' rate requirements by iteratively addressing the optimizations of the UAV's locations and of the boundaries of their coverage areas. The authors in \cite{Mozaffari16Arxiv2} analyze the downlink coverage and rate performance for static and mobile UAV. For a static UAV, they derive coverage probability and system sum-rate as a function of the UAV's altitude and of the number of users. For a mobile UAV, the minimum number of stop points for the UAV required to completely cover the area of interest is analyzed via disk covering problem. A point-to-point communication link between the UAV and a ground user is investigated in \cite{Zeng16Arxiv2} with the goal of optimizing the UAV's trajectory under a UAV's energy consumption model that accounts for the impact of the UAV's velocity and acceleration.

Beside the communication scenarios reviewed above, other optimization problems involving UAV path planning have been studied. For instance, in \cite{Manyam16Arxiv}, a scenario is investigated in which a ground vehicle and an aerial vehicle move cooperatively to carry out intelligence, surveillance and reconnaissance (ISR) missions. Path planning for the ground and aerial vehicles is carried out via a branch-and-cut algorithm. As another example, reference \cite{Dorling16TSMC} tackles the problem of UAV trajectory optimization for drone delivery of material goods by minimizing the total energy cost under a delivery time limit constraint, as well as by minimizing the overall delivery time under a energy budget constraint. Sub-optimal solutions for the problem of interest are presented via a simulated annealing heuristic approach. 

\subsection{UAV as a Moving Cloudlet}\label{sec:uav}     
As briefly reviewed above, most prior works on the deployment of UAVs in communication system assume their use either as moving relays \cite{Zeng16TCOM, Zhao04ACM, Shah03Adhoc, Zhan11TAES} or as flying base stations \cite{Mozaffari15Arxiv, Mozaffari16Arxiv, Mozaffari16Arxiv2, Soorki16Arxiv, Zeng16Arxiv2, Manyam16Arxiv, Dorling16TSMC}. It was instead noted in \cite{Loke15Arxiv} that UAVs can also be used as mobile cloud computing systems, in which a UAV-mounted cloudlet \cite{Zeng16Comm, Loke15Arxiv, JSA17IET} provides application offloading opportunities to mobile users (MUs). UAVs can hence enable fog computing \cite{Bonomi14} even in the absence of a working wireless infrastructure. Specifically, MUs can offload computationally heavy tasks, such as object recognition or augmented-reality applications, to the cloudlet by means of uplink/downlink communications with the UAV. Referring to Fig. \ref{fig:sys} for an illustration, the offloading procedure requires uplink transmission of input data for the application to be run at the cloudlet from the mobiles to the UAV, computing at the UAV-mounted cloudlet, and downlink transmission of outcome of computing at the cloudlet from the UAV to the mobiles. Among the possible examples and applications, the use of the moving cloudlets can for instance play an important role in disaster response, emergency relief or military scenarios, as mobile devices with limited processing capabilities can benefit from the cloudlet-aided execution of data analytics application for the assessment of the status of victims, enemies, or hazardous terrain and structures.

\begin{table}[t]
\caption{List of symbols} \label{t1}
\begin{center}
    \begin{tabular}{|p{1.5cm}|p{6.5cm}|}
    \hline
    Parameter & Definition \\ \hline\hline
    $K$ & Number of mobile users (MUs) \\ \hline
    $I_k$ & Number of input information bits of MU $k$ to be processed \\ \hline
    $C_k$ & Number of CPU cycles per input bit of MU $k$ needed for computing \\ \hline
    $O_k$ & Number of output bits produced by the execution of the application per input bits of MU $k$ \\ \hline
    $T$ & Latency constraint or deadline \\ \hline
    $N$ & Number of frames within $T$ \\ \hline
    $\pmb{p}^\m_k$  & Position of MU $k$ \\ \hline
    $\pmb{p}^\cc(t)$ ($\pmb{p}^\cc_n$) & Position of UAV \\ \hline
    $\pmb{p}^\cc_I$ ($\pmb{p}^\cc_{1}$) & Initial position of UAV projected onto xy-plane \\ \hline 
    $\pmb{p}^\cc_F$ ($\pmb{p}^\cc_{N+1}$) & Final position of UAV projected onto xy-plane \\ \hline    
    $H$ & Altitude of the UAV \\ \hline
    $\pmb{v}^\cc_n$ & UAV's velocity at the $n$th frame \\ \hline
    $\pmb{v}^\cc$ & UAV's initial and final velocity constraint \\ \hline 
    $v_{\max}$ & UAV's maximum speed \\ \hline
    $\pmb{a}^\cc_n$ & UAV's acceleration at the $n$th frame \\ \hline
    $a_{\max}$ & UAV's maximum acceleration \\ \hline
    $\Delta$ & Frame duration \\ \hline
    $\mathcal{E}$ & UAV's energy budget \\ \hline    
    $\pmb{g}_{k,n}(\pmb{p}^\cc_n)$ & Path loss between MU $k$ and cloudlet at the $n$th frame \\ \hline
    $g_0$ & Received power at the reference distance $d_0 = 1$ m for a transmission power of $1$ W \\ \hline
    $E^\m$ & Total energy consumption in mobile execution \\ \hline 
    $E^\m_k$ & Energy consumption of MU $k$ in mobile execution \\ \hline
    $E^\cc_{k,n}$ & Computation energy consumption at cloudlet for MU $k$ at the $n$th frame \\ \hline   
    $E^d_{O,k,n}$ & Transmission energy consumption for communication between MU $k$ and cloudlet at the $n$th frame in orthogonal access ($d=\m$ for uplink, $d=\cc$ for downlink) \\ \hline
    $E^d_{N,k,n}$ & Transmission energy consumption for communication between MU $k$ and cloudlet at the $n$th frame in non-orthogonal access ($d=\m$ for uplink, $d=\cc$ for downlink)  \\ \hline 
    $E_{F,n}^\cc$ & Flying energy consumption of the $n$th frame  \\ \hline     
    $L^d_{k,n}$ & Number of bits transmitted for communication between MU $k$ and cloudlet at the $n$th frame ($d=\m$ for uplink, $d=\cc$ for downlink) \\ \hline
    $l_{k,n}$ & Number of bits computed for application of MU $k$ at cloudlet in $n$th frame  \\ \hline
    $f^\m_k$ & CPU frequency of MU $k$  \\ \hline
    $f^\cc_n$ & CPU frequency of cloudlet at the $n$th frame  \\ \hline
    $B$ & Bandwidth \\ \hline
    $N_0$ & Noise spectrum density \\ \hline
    $\gamma_k^\m$ & Effective switched capacitance of MU $k$'s processor \\ \hline
    $\gamma^\cc$ & Effective switched capacitance of cloudlet processor \\ \hline
    $M$ & UAV's gross mass  \\ \hline
    $g$ & Gravitational acceleration  \\ \hline
    $\kappa$ & Constant for Model 1 in (\ref{eq:pro_simple}) ($\kappa = 0.5 M\Delta$) \\ \hline
    $\kappa_1$ & Constant for Model 2 in (\ref{eq:pro}) ($\kappa_1=0.5\rho C_{D_0} S_r\Delta$ for fixed-wing UAV and $\kappa_1=0.5\rho C_{D_f} S_r\Delta$ for rotary-wing UAV) \\ \hline
    $\kappa_2$ & Constant for Model 2 in (\ref{eq:pro}) ($\kappa_2=2 M^2g^2\Delta/(\pi e_0 A_R \rho S_r)$ for fixed-wing UAV and $\kappa_2=\epsilon M^2g^2\Delta/(2 \rho A)$ for rotary-wing UAV) \\ \hline
    \end{tabular}
    \end{center}
\end{table} 
\subsection{Main Contributions}\label{sec:contri}     
In this paper, we focus on the scenario illustrated in Fig. \ref{fig:sys} in which a moving UAV is deployed to offer offloading opportunities to mobile devices. We tackle the key design problem of optimizing the bit allocation for communication in uplink and downlink and for computing at the cloudlet, as well as the UAV's trajectory, with the goal of minimizing the mobile energy consumption. For uplink and downlink transmission, we assume frequency division duplex (FDD) and either orthogonal or non-orthogonal multiple access (NOMA) schemes. We note that the latter is a promising multiple access technique for 5G networks which is currently being considered due to its potentially superior spectral efficiency \cite{Saito13PIMRC, Ding14SPL}. The design problem is formulated for both orthogonal access and non-orthogonal access under latency and UAV's energy budget constraints. The UAV's energy budget includes the energy consumption for communication and computing as well as for flying. For the latter energy constraint, we consider two different models, both of which are investigated in the literature. The first model, adopted in \cite{Xue14, Borst08JOA, Chakrabarty09GNCC, Chakrabarty11JGCD}, postulates the flying energy to depend only on the UAV's velocity, while the second model accounts also for the impact of the acceleration following \cite{Zeng16Arxiv2, Filippone06, Leishman06, Kong09AIAA}. The resulting non-convex problem is tackled by means of successive convex approximation (SCA) \cite{Scutari14Arxiv, Scutari16ArxivII}, which allows us to derive an efficient iterative algorithm that is guaranteed to converge to a local minimum of the original non-convex problem.  

The rest of this paper is organized as follows. Section \ref{sec:sys} presents the system model including the energy consumption models for communication, computation and flying. In Section \ref{sec:CE_oma} and Section \ref{sec:CE_noma}, we formulate and tackle the mentioned joint optimization problems over the bit allocation and UAV's trajectory under the first UAV's flying energy consumption model for orthogonal access and NOMA, respectively. Then, in Section \ref{sec:pro}, the joint optimization problems are studied with the second UAV's flying energy consumption model. Finally, numerical results are given in Section \ref{sec:num}, and conclusions are drawn in Section \ref{sec:con}.
\section{System Model}\label{sec:sys}
\subsection{Set-Up}\label{sec:setup}
\begin{figure}[t]
\begin{center}
\includegraphics[width=8cm]{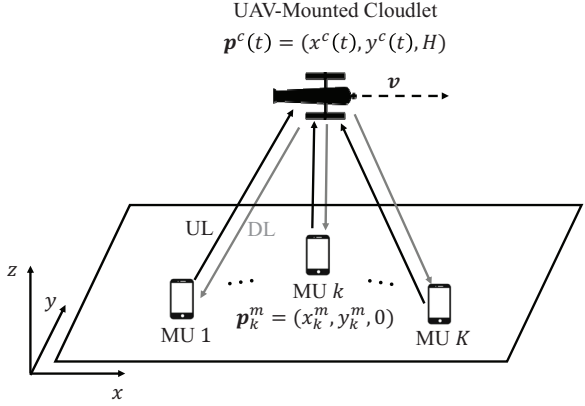}
\caption{Illustration of the considered mobile cloud computing system based on a UAV-mounted cloudlet that provides application offloading opportunities to MUs. The key design problem is the optimization of the bit allocation for communication in uplink (UL) and downlink (DL) and computing, as well as the cloudlet's trajectory with the goal of minimizing the mobile energy consumption.} \label{fig:sys}
\end{center}
\end{figure}

In this paper, we consider the mobile cloud computing system illustrated in Fig. \ref{fig:sys}, which consists of $K$ MUs and a UAV-mounted cloudlet. We study the optimization of the offloading process from the MUs to the moving cloudlet with the goal of minimizing the total energy consumption of all the MUs. To enable the offloading of a given application for each MU $k$, with $k \in \mathcal{K}=\{1, \dots, K\}$, the following steps are necessary; $(i)$ uplink transmission of the application input data from the MU $k$ to the UAV; $(ii)$ execution of the application by the UAV-mounted cloudlet; and $(iii)$ downlink transmission of the output of the application from the UAV to MU $k$. We assume frequency division duplex (FDD) with equal channel bandwidth $B$ allocated for uplink and downlink. Moreover, for uplink and downlink communications, two types of access schemes are considered, namely orthogonal and non-orthogonal access. We note that, in $5$G, the latter is typically referred to as NOMA. Receivers at the MUs and cloudlet are assumed to have no limitations on the resolution of their digital front-ends. The application of the MU $k \in \mathcal{K}$ is characterized by the number $I_k$ of input information bits to be processed, the number $C_k$ of CPU cycles per input bit needed for computing, and the number $O_k$ of output bits produced per input bit by the execution of the application. We assume that all applications need to be computed within a time $T$.

\begin{figure}[t]
\begin{center}
\includegraphics[width=9cm]{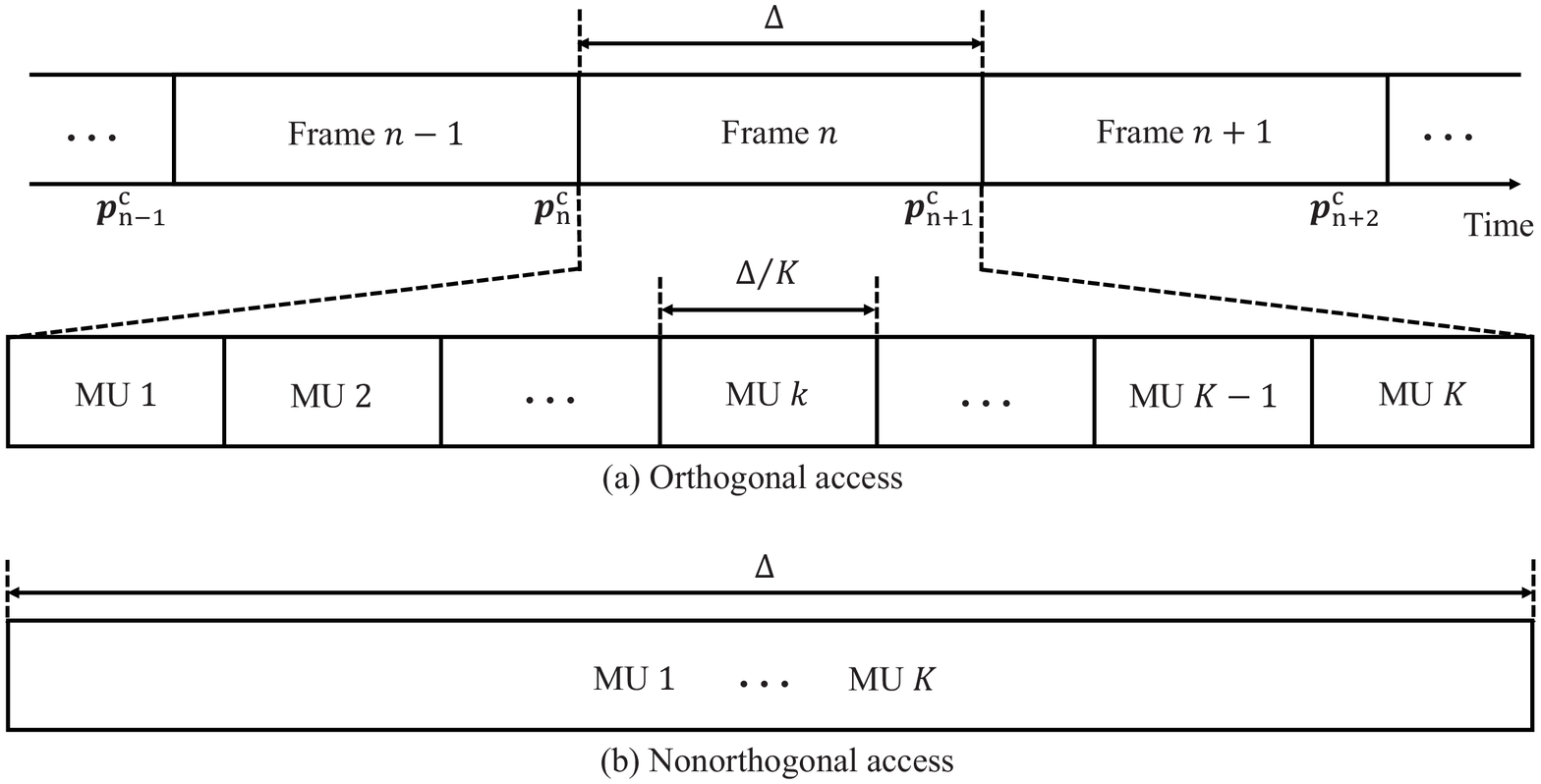}
\caption{Frame structure of the considered mobile cloud computing system: (a) Orthogonal access, (b) Non-orthogonal access.} \label{fig:frame}
\end{center}
\end{figure}

A three-dimensional Cartesian coordinate system is adopted, as shown in Fig. \ref{fig:sys}, whose coordinates are measured in meters. We assume that all MUs are located at the $xy$-plane, e.g., on the ground, with MU $k$ located at position $\pmb{p}^\m_k=(x^\m_k,y^\m_k,0)$, for $k \in \mathcal{K}$, while the UAV flies along a trajectory $\pmb{p}^{\cc}(t)=(x^\cc(t),y^\cc(t),H)$ with a fixed altitude $H$, for $0 \le t \le T$. In this work, since the UAV flies horizontally at a constant altitude $H$, we focus on the UAV's trajectory projected onto the xy-plane. Due to its launching and landing locations, flying paths and operational capability, the initial and final location and maximum speed of the UAV are assumed to be predetermined as $\pmb{p}^\cc_I=(x^\cc_I, y^\cc_I)$, $\pmb{p}^\cc_F=(x^\cc_F, y^\cc_F)$, both with the altitude $H$, and $v_{\max}$, respectively. 

As seen in Fig. \ref{fig:frame}, the time horizon $T$ is divided into $N$ intervals each of duration $\Delta$ seconds \cite{JSA17IET, Zeng16TCOM, Zeng16Arxiv2}, i.e., $T=N\Delta$, in which the UAV continuously communicates and computes while flying. The frame duration $\Delta$ is chosen to be sufficiently small for the UAV's location to be approximately constant within each frame. Accordingly, the UAV's trajectory $\pmb{p}^{\cc}(t)$ can be characterized by the discrete-time UAV's location $\pmb{p}^\cc_n = (x^\cc_n, y^\cc_n)$ with altitude $H$, for $n \in \mathcal{N}=\{1, \dots, N\}$, where $\pmb{p}_1^\cc=\pmb{p}_I^\cc$ and $\pmb{p}_{N+1}^\cc=\pmb{p}_F^\cc$. The trajectory $\{\pmb{p}^\cc_n\}_{n \in \{2, \dots, N\}}$ is subject to optimization. The quantity 
\begin{equation}\label{eq:v}
\pmb{v}_n^\cc = \frac{\pmb{p}_{n+1}^\cc-\pmb{p}_{n}^\cc}{\Delta}
\end{equation}
represents the velocity vector in the $n$th frame. As mentioned, we have the constraint on the maximum speed
\begin{equation}\label{eq:vmax}
\|\pmb{v}_n^\cc\| = \frac{\left\|\pmb{p}_{n+1}^\cc-\pmb{p}_{n}^\cc\right\|}{\Delta} \le v_{\max}.
\end{equation}
Note that the final position should be assumed no later than after a time $T$ from the initial time. As a result, we have the constraint
\begin{equation}\label{eq:vmax_tot}
\frac{\left\|\pmb{p}_{N+1}^\cc-\pmb{p}_{1}^\cc\right\|}{N\Delta} \le v_{\max},
\end{equation} 
in order for a feasible trajectory from the UAV's initial to final location to exist. 

For \textit{orthogonal access}, each $n$th frame, for $n \in \mathcal{N}$, is assumed to have $K$ equally spaced time slots, each of which has the duration of $\Delta/K$ seconds and is preallocated to one MU in both uplink and downlink. For \textit{non-orthogonal access}, all MUs simultaneously transmit and receive data within the entire frame of $\Delta$ seconds in uplink and downlink. In the latter case, we treat the interference from undesired signals as additive noise. This assumption is standard in the practical implementation of communication systems, as well as in the communication and information theory literatures (see, e.g., \cite{Geng15TIT}). We recall that uplink and downlink do not interfere with one another due to the assumption of FDD.             

As in \cite{JSA17IET, Zeng16TCOM, Zeng16Arxiv2}, we assume that the communication channels between MUs and UAV are dominated by line-of-sight links. At the $n$th frame, the channel gain between the MU $k$ and cloudlet is accordingly given by \cite{JSA17IET, Zeng16TCOM, Zeng16Arxiv2} 
\begin{equation}\label{eq:ch}
g_{k,n}(\pmb{p}^\cc_{n})=\frac{g_0}{\left(x^\cc_{n}-x^\m_k\right)^2+\left(y^\cc_{n}-y^\m_k\right)^2+H^2},
\end{equation} 
where $g_0$ represents the received power at the reference distance $d_0=1$ m for a transmission power of $1$ W. An additive white Gaussian channel noise with zero mean and power spectral density $N_0$ [dBm/Hz] is assumed. In the following, we summarize the energy consumption model for computation \cite{Yuan03ACM, Yuan06ACM}, communication \cite{JSA17IET, Zeng16TCOM, Zeng16Arxiv2} and flying \cite{Xue14, Borst08JOA, Chakrabarty09GNCC, Chakrabarty11JGCD}. As we will detail in the following sections, our goal is to minimize the mobile energy consumption. 

\subsection{Energy Consumption Model for Offloading}\label{sec:energy}
\textit{Computation energy}: First, we review the energy consumption model for computation at the cloudlet \cite{Yuan03ACM, Yuan06ACM}. When the CPU of the cloudlet is operated at the frequency $f^\cc$ [CPU cycles/s], the energy consumption required for executing the application of MU $k$ over $l$ input bits is given as 
\begin{equation}\label{eq:Ecomp}
E^\cc_k(l, f^\cc) = \gamma^\cc C_k l(f^\cc)^2,
\end{equation}
where $\gamma^\cc$ is the effective switched capacitance of the cloudlet processor.

\textit{Communication energy}: The energy consumption for communication at the mobile and at the UAV depends on whether orthogonal access or non-orthogonal access are deployed. With \textit{orthogonal access}, the energy consumption for transmitting $L^\m_{k,n}$ bits in the uplink, or $L^\cc_{k,n}$ in the downlink, between the MU $k$ and cloudlet, within the allocated slot $\Delta/K$ seconds at the $n$th frame, can be computed based on standard information-theoretic arguments \cite{Cov06} as
\begin{equation}\label{eq:Ecomm_oma}
E_{\oo,k,n}^d(L^d_{k,n}, \pmb{p}^\cc_n)=\frac{N_0 B \Delta/K}{g_{k,n}(\pmb{p}^\cc_n)}\left(2^{\frac{L^d_{k,n}}{B\Delta/K}}-1\right),
\end{equation}
where we recall that $g_{k,n}(\pmb{p}^\cc_n)$ in (\ref{eq:ch}) is the path loss between the MU $k$ and cloudlet at the $n$th frame, and $d = \m$ for uplink while $d = \cc$ for downlink.

With \textit{non-orthogonal access}, e.g., NOMA in 5G, since all the MUs can simultaneously transmit and receive data within entire frame of duration $\Delta$ in both uplink and downlink, interference is caused by the undesired signals of other MUs which are assumed to be treated as additive noise \cite{Geng15TIT}. When $L^\m_{k,n}$ and $L^\cc_{k,n}$ bits are transmitted in uplink and in downlink, respectively, between the MU $k$ and cloudlet experiencing a path loss $g_{k,n}(\pmb{p}^\cc_n)$ at the $n$th frame, the transmission energy consumptions of uplink and downlink are calculated as \cite{Cov06}
\begin{subequations}
\begin{eqnarray}
&&\hspace{-0.8cm} E_{\nn,k,n}^{\m}(L^\m_n, \pmb{p}^\cc_n) = \frac{1}{g_{k,n}(\pmb{p}^\cc_n)}\left(N_0 B\Delta \right.\nonumber\\
&&\hspace{-1cm} \left.+ \sum_{k'=1, k'\neq k}^{K} g_{k',n}(\pmb{p}^\cc_n)E_{\nn,k',n}^{\m}(L_n^{\m}, \pmb{p}^\cc_n)\right)\left(2^{\frac{L^{\m}_{k,n}}{B\Delta}}-1\right)\label{eq:Ecomm_noma_up}\\
&&\hspace{-1.5cm} \text{and}\hspace{0.2cm}  E_{\nn,k,n}^{\cc}(L^\cc_n, \pmb{p}^\cc_n) = \left(\frac{N_0 B\Delta}{g_{k,n}(\pmb{p}^\cc_n)} \right.\nonumber\\
&&\hspace{-1cm} \left. + \sum_{k'=1, k'\neq k}^{K} E_{\nn,k',n}^{\cc}(L^{\cc}_n, \pmb{p}^\cc_n)\right)\left(2^{\frac{L^{\cc}_{k,n}}{B\Delta}}-1\right),\label{eq:Ecomm_noma_down}
\end{eqnarray}
\end{subequations}
respectively, where the sets of all the uplink and downlink transmission bits related to the $n$th frame are denoted as $L^\m_n=\{L^\m_{k,n}\}_{k \in \mathcal{K}}$ and $L^\cc_n=\{L^\cc_{k,n}\}_{k \in \mathcal{K}}$. Note that in the non-orthogonal access, the transmission energies required for the applications of MU $k \in \mathcal{K}$ in both uplink and downlink depend on the transmission energies of the other MUs due to the interference.  

\textit{Flying energy}: As for the energy consumption at the UAV due to flying, we will consider two different models that have been adopted in the literature. The first model considered in, e.g., \cite{Xue14, Borst08JOA, Chakrabarty09GNCC, Chakrabarty11JGCD}, postulates the flying energy at each frame $n$ to depend only on the velocity vector $\pmb{v}_n^\cc$ as
\begin{equation}\label{eq:pro_simple}
\hspace{-0.5cm}\text{(Model 1)}\hspace{0.2cm}E_{F,n}^\cc(\pmb{v}_n^\cc) = \kappa\left\|\pmb{v}_n^\cc\right\|^2,
\end{equation}
where $\kappa=0.5M\Delta$ and $M$ is the UAV's mass, including its payload. Note that only the kinetic energy is accounted for in Model 1, due to the fact that constant-height flight entails no change in the gravitational potential energy. The second model assumes that the energy $E_{F,n}^\cc$ depends also on the acceleration vector (cf. (\ref{eq:pro})) according to \cite{Zeng16Arxiv2, Filippone06, Leishman06, Kong09AIAA}. We will describe and study this model in Section \ref{sec:pro}.     

\subsection{Energy Consumption Model for Mobile Execution}\label{sec:ME}
For reference, we consider the total energy consumption of the MUs if all applications are executed locally. In order to guarantee that each MU $k$ processes the $I_k$ input bits within $T$ seconds, the CPU frequency $f^\m_k$ must be chosen as \cite{Yuan03ACM, Yuan06ACM}  
\begin{equation}\label{eq:optf_ME}
f^\m_k = \frac{C_kI_k}{T}, 
\end{equation} 
which yields the total energy consumption of MUs of  
\begin{equation}\label{eq:optE_ME_tot} 
E^{\m} \triangleq \sum_{k=1}^{K}E^\m_k(I_k, f^\m_k) = \sum_{k=1}^{K}\frac{\gamma^\m_k C_k^3}{T^2}I_k^3,
\end{equation}
where $\gamma^\m_k$ is the effective switched capacitance of the MU $k$'s processor. 

\section{Optimal Energy Consumption for Orthogonal Access}\label{sec:CE_oma}
In this section, we tackle the problem of minimizing the total mobile energy consumption for offloading assuming orthogonal access in uplink and downlink. Specifically, we focus on the joint optimization of the bit allocation for uplink and downlink data transmission and for cloudlet's computing, as well as of the cloudlet's trajectory, under constraints on the UAV's energy budget and mobility constraints. We consider the model (\ref{eq:pro_simple}) for the UAV flying model.  

\subsection{Problem Formulation}\label{sec:prob_oma}
At the $n$th frame, for $n \in \mathcal{N}$, we define the number of input bits transmitted in the uplink from the MU $k$ to cloudlet as $L_{k,n}^\m$, the number of bits computed for the application of the MU $k$ at the cloudlet as $l_{k,n}$, and the number of bits transmitted in the downlink from cloudlet to MU $k$ as $L_{k,n}^\cc$. Also, we denote the frequency at which the cloudlet CPU is operated for the offloaded applications from MUs at the $n$th frame as $f^{\cc}_n$. Along with the cloudlet position $\{\pmb{p}^\cc_n\}$, these variables are subject to optimization.  

According to the definitions above, at every $n$th frame, the CPU frequency $f^{\cc}_n$ selected by the UAV must be such that the UAV can process $\sum_{k=1}^{K}l_{k,n}$ bits from the applications of all the MUs within the given frame as
\begin{equation}\label{eq:optf_CE}
f^{\cc}_n=\frac{\sum_{k=1}^{K}C_kl_{k,n}}{\Delta}. 
\end{equation}   
This yields the computation energy required for offloading by MU $k$ at the $n$th frame as 
\begin{equation}\label{eq:optEk_CE}
E^\cc_{k,n}(l_n) \triangleq E^\cc_k(l_n, f_n^{\cc})= \frac{\gamma^\cc C_k l_{k,n}}{\Delta^2}\left(\sum_{k'=1}^{K}C_{k'}l_{k',n}\right)^2,
\end{equation}    
where we have defined the total number of computing bits at the $n$th frame as $l_n=\{l_{k,n}\}_{k \in \mathcal{K}}$. Our objective is to minimize the total energy consumption at the MUs by jointly optimizing the bit allocation $\{L_{k,n}^\m\}_{n \in \{1, \dots, N-2\}, k \in \mathcal{K}}$, $\{l_{k,n}\}_{n \in \{2, \dots, N-1\}, k \in \mathcal{K}}$ and $\{L_{k,n}^\cc\}_{n \in \{3, \dots, N\}, k \in \mathcal{K}}$ for communication and computing needed to support offloading from all MUs along with the cloudlet trajectory $\{\pmb{p}^\cc_n\}_{n \in \{2, \dots, N\}}$. The corresponding design problem is formulated as follows:
\begin{subequations}\label{eq:CEopt}
\begin{eqnarray}
&& \hspace{-1.5cm} \underset{ \{L_{k,n}^\m\}, \{l_{k,n}\}, \{L_{k,n}^\cc\}, \{\pmb{p}^\cc_n\}}{\text{minimize}} \hspace{0.3cm} \sum_{k=1}^{K}\sum_{n=1}^{N-2} E_{\oo,k,n}^\m(L_{k,n}^\m, \pmb{p}^\cc_n)\label{eq:obj}\\
&& \hspace{-1.4cm}  \text{s.t.} \hspace{0.2cm} \sum_{k=1}^{K}\sum_{n=1}^{N-2}  E^\cc_{k,n+1}(l_{n+1}) + E_{\oo,k,n+2}^\cc(L_{k,n+2}^\cc, \pmb{p}^\cc_{n+2}) \nonumber\\
&& \hspace{1cm} +  \sum_{n=1}^N E_{F,n}^\cc(\pmb{v}_n^\cc) \le \mathcal{E}\label{eq:lifetime}\\
&& \hspace{-0.7cm}  \sum_{i=1}^{n}l_{k,i+1} \le \sum_{i=1}^{n}L_{k,i}^\m,\nonumber\\
&& \hspace{1cm} \text{for} \hspace{0.2cm} k \in \mathcal{K} \hspace{0.2cm} \text{and} \hspace{0.2cm} n=1, \dots, N-2 \label{eq:ineq_mc}\\
&& \hspace{-0.7cm}  \sum_{i=1}^{n}L_{k,i+2}^\cc \le O_k \sum_{i=1}^{n}  l_{k,i+1}, \nonumber\\ 
&& \hspace{1cm} \text{for} \hspace{0.2cm} k \in \mathcal{K} \hspace{0.2cm} \text{and} \hspace{0.2cm} n=1, \dots, N-2 \label{eq:ineq_cm}\\
&& \hspace{-0.7cm} \sum_{n=1}^{N-2} L_{k,n}^\m=I_k, \hspace{0.2cm} \text{for} \hspace{0.2cm} k \in \mathcal{K} \label{eq:eq_Lmc}\\
&& \hspace{-0.7cm} \sum_{n=1}^{N-2} l_{k,n+1}=I_k, \hspace{0.2cm} \text{for} \hspace{0.2cm} k \in \mathcal{K}  \label{eq:eq_lc}\\
&& \hspace{-0.7cm} \sum_{n=1}^{N-2} L_{k,n+2}^\cc=O_k I_k, \hspace{0.2cm} \text{for} \hspace{0.2cm} k \in \mathcal{K} \label{eq:eq_Lcm}\\
&& \hspace{-0.7cm} L_{k,n}^\m, l_{k,n}, L_{k,n}^\cc \ge 0, \hspace{0.2cm} \text{for} \hspace{0.2cm} k \in \mathcal{K} \hspace{0.2cm} \text{and} \hspace{0.2cm} n \in \mathcal{N} \label{eq:pos_oma}\\
&& \hspace{-0.7cm} \pmb{p}^\cc_1 = \pmb{p}^\cc_I,  \pmb{p}^\cc_{N+1} = \pmb{p}^\cc_F, \hspace{0.2cm} \label{eq:eq_pos} \\
&& \hspace{-0.7cm} \left\|\pmb{v}^\cc_n\right\| \le v_{\max}, \hspace{0.2cm} \text{for} \hspace{0.2cm} n \in \mathcal{N} \label{eq:ineq_v}\\
&& \hspace{-0.7cm}  \pmb{v}_n^\cc = \frac{\pmb{p}_{n+1}^\cc-\pmb{p}_{n}^\cc}{\Delta} \hspace{0.2cm} \text{for} \hspace{0.2cm} n \in \mathcal{N}, \label{eq:eq_v}
\end{eqnarray}
\end{subequations}  
where $\pmb{v}_n^\cc$ is defined in (\ref{eq:eq_v}) (cf. (\ref{eq:v})); the energies $E_{\oo,k,n}^\m(\cdot)$ and $E_{\oo,k,n}^\cc(\cdot)$ needed for uplink and downlink communication between MU $k$ and cloudlet in (\ref{eq:obj}) and (\ref{eq:lifetime}), respectively, are defined in (\ref{eq:Ecomm_oma}); and $\mathcal{E}$ in (\ref{eq:lifetime}) represents the UAV energy budget constraint, accounting for offloading and flying. In problem (\ref{eq:CEopt}), the inequality constraints (\ref{eq:ineq_mc}) and (\ref{eq:ineq_cm}) ensure that the number of bits computed at the $(n+1)$th frame by the cloudlet is no larger than the number of bits received by the cloudlet in the uplink in the previous $n$ frames, and the number of bits transmitted from the cloudlet in the downlink at the $(n+2)$th frame is no larger than the number of bits available at the cloudlet after computing in the previous $(n+1)$ frames, respectively, for the MU $k \in \mathcal{K}$ and $n=1, \dots, N-2$. The equality constraints (\ref{eq:eq_Lmc}) - (\ref{eq:eq_Lcm}) enforce the completion of offloading while (\ref{eq:pos_oma}) is imposed for the non-negative bit allocations. The constraints (\ref{eq:eq_pos}) and (\ref{eq:ineq_v}) guarantee the cloudlet's initial and final position constraint and maximum speed constraints, respectively. 

\subsection{Successive Convex Approximation}\label{sec:SCA_oma}
The problem (\ref{eq:CEopt}) is non-convex due to the non-convex objective function (\ref{eq:obj}) and non-convex constraint (\ref{eq:lifetime}). To tackle this problem without resorting to expensive global optimization methods, we develop an SCA-based algorithm that builds on the inner convex approximation framework proposed in \cite{Scutari14Arxiv, Scutari16ArxivII}. This approach prescribes the iterative solution of problems in which the non-convex objective function and constraints are replaced by suitable convex approximations. Each problem can be further solved in a distributed manner by using dual decomposition techniques.  

In order to develop the SCA-based algorithm, we use the following lemmas. 
\begin{lemma} 
([29, Example 8]) Given a non-convex objective function $U(\pmb{x})=f_1(\pmb{x})f_2(\pmb{x})$, with $f_1$ and $f_2$ convex and non-negative, for any $\pmb{y}$ in the domain of $U(\pmb{x})$, a convex approximant of $U(\pmb{x})$ that has the properties required by the SCA algorithm \cite[Assumption 2]{Scutari14Arxiv} is given as  
\begin{equation}\label{eq:obj_approx}
\bar{U}(\pmb{x}; \pmb{y}) \triangleq f_1(\pmb{x})f_2(\pmb{y}) + f_1(\pmb{y})f_2(\pmb{x}) + \frac{\tau_i}{2}(\pmb{x}-\pmb{y})^T\pmb{H}(\pmb{y})(\pmb{x}-\pmb{y}),
\end{equation}  
where $\tau_i >0$ is a positive constant (ensuring that (\ref{eq:obj_approx}) is strongly convex) and  $\pmb{H}(\pmb{y})$ is a positive definite matrix. 
\end{lemma}

\begin{lemma}
([29, Example 4]) Given a non-convex constraint $g(\pmb{x}_1, \pmb{x}_2) \le 0$, where $g(\pmb{x}_1, \pmb{x}_2) = h_1(\pmb{x}_1)h_2(\pmb{x}_2)$ is the product of $h_1$ and $h_2$ convex and non-negative, for any $(\pmb{y}_1, \pmb{y}_2)$ in the domain of $g(\pmb{x}_1, \pmb{x}_2)$, a convex approximation that satisfies the conditions \cite[Assumption 3]{Scutari14Arxiv} required by the SCA algorithm is given as 
\begin{eqnarray}\label{eq:con_approx}
&&\hspace{-1cm} \bar{g}(\pmb{x}_1, \pmb{x}_2; \pmb{y}_1, \pmb{y}_2) \nonumber\\
&&\hspace{-1cm} \triangleq \frac{1}{2}\left(h_1(\pmb{x}_1)+ h_2(\pmb{x}_2)\right)^2-\frac{1}{2}\left(h_1^2(\pmb{y}_1)+ h_2^2(\pmb{y}_2)\right) - \nonumber\\
&&\hspace{-1cm} h_1(\pmb{y}_1)h_1^{'}(\pmb{y}_1)(\pmb{x}_1-\pmb{y}_1) - h_2(\pmb{y}_1)h_2^{'}(\pmb{y}_2)(\pmb{x}_2-\pmb{y}_2).\nonumber\\
\end{eqnarray}
\end{lemma}

We recall that, beside technical conditions on continuity and smoothness, the SCA algorithm requires the strongly convex approximation of the objective function to have the same first derivative of the objective function, while the convex approximation of the constraints is required to be tight at the approximation point and to upper bound the original constraints. 

To proceed, define the set of primal variables for problem (\ref{eq:CEopt}) as $\pmb{z}=\{\pmb{z}_n\}_{n \in \mathcal{N}}$ with $\pmb{z}_n=(\{L_{k,n}^\m\}_{k \in \mathcal{K}},$ $\{l_{k,n}\}_{k \in \mathcal{K}},\{L_{k,n}^\cc\}_{k \in \mathcal{K}}, \pmb{p}^\cc_n)$ being the optimization variables for the $n$th frame. We observe that the function $E_{\oo,k,n}^\m(\pmb{z}_n) \triangleq E_{\oo,k,n}^\m(L_{k,n}^\m, \pmb{p}^\cc_n)$ is the product of two convex and non-negative functions, namely
\begin{subequations}
\begin{eqnarray}
&& \hspace{-1.5cm} f_1(L_{k,n}^\m)=\frac{N_0 B \Delta/K}{g_0} \left(2^{\frac{L_{k,n}^\m}{B\Delta/K}}-1\right)\\
&& \hspace{-2.5cm} \text{and} \hspace{0.5cm} f_2(\pmb{p}^\cc_n) = \left(x^\cc_n - x^\m_k\right)^2+\left(y^\cc_n - y^\m_k\right)^2+H^2.
\end{eqnarray}
\end{subequations}
Then, using Lemma 1 and defining $\pmb{z}_n(v)=(\{L_{k,n}^\m(v)\}_{k \in \mathcal{K}}, \{l_{k,n}(v)\}_{k \in \mathcal{K}}, \{L_{k,n}^\cc(v)\}_{k \in \mathcal{K}}, \pmb{p}^\cc_n(v)) \in \mathcal{X}$ for the $v$th iterate within the the feasible set $\mathcal{X}$ of (\ref{eq:CEopt}), we obtain a strongly convex surrogate function $\bar{E}_{\oo,k,n}^\m(\pmb{z}_n; \pmb{z}_n(v))$ of $E_{\oo,k,n}^\m(\pmb{z}_n)$ as
\begin{eqnarray}\label{eq:commE_approx_oma}
&&\hspace{-0.7cm} \bar{E}_{\oo,k,n}^\m(\pmb{z}_n; \pmb{z}_n(v))\triangleq\bar{E}_{\oo,k,n}^\m(L_{k,n}^\m, \pmb{p}^\cc_n; L_{k,n}^\m(v), \pmb{p}^\cc_ n(v)) \nonumber\\
&&\hspace{-0.7cm} = f_1(L_{k,n}^\m)f_2(\pmb{p}^\cc_n(v)) + f_1(L_{k,n}^\m(v))f_2(\pmb{p}^\cc_n) \nonumber\\
&&\hspace{-0.3cm} +\frac{\tau_{L_{k,n}^\m}}{2}\left(L_{k,n}^\m-L_{k,n}^\m(v)\right)^2+\frac{\tau_{x^\cc_n}}{2}\left(x^\cc_n-x^\cc_n(v)\right)^2\nonumber\\
&&\hspace{-0.3cm} +\frac{\tau_{y^\cc_n}}{2}\left(y^\cc_n-y^\cc_n(v)\right)^2,
\end{eqnarray}     
where $\tau_{L_{k,n}^\m}, \tau_{x^\cc_n}, \tau_{y^\cc_n} > 0$.  

For the non-convex constraint (\ref{eq:lifetime}), we derive a convex upper bound using Lemma 2 given that the constraint can be written as the sum of two products of convex functions, namely 
\begin{subequations}
\begin{eqnarray}
&& \hspace{-1.8cm} E^\cc_{k,n}(\pmb{z}_n) \triangleq E^\cc_{k,n}(l_{n})=\frac{\gamma^\cc C_k}{\Delta^2}g(\pmb{x}_1, \pmb{x}_2) \label{eq:re_comp}\\
&&\hspace{-2.8cm} \text{and}\hspace{0.5cm} E_{\oo,k,n}^\cc(\pmb{z}_n) \triangleq E_{\oo,k,n}^\cc(L_{k,n}^\cc, \pmb{p}^\cc_n)\nonumber\\
&&\hspace{-0.1cm} =\frac{N_0 B \Delta/K}{g_0}g(\pmb{x}_1, \pmb{x}_2), \label{eq:re_comm}
\end{eqnarray}
\end{subequations} 
where $h_1(\pmb{x}_1)=l_{k,n}$ and $h_2(\pmb{x}_2)=(\sum_{k'=1}^{K}C_{k'}l_{k',n})^2$ with $\pmb{x}_1=l_{k,n}$ and $\pmb{x}_2=l_n=\{l_{k',n}\}_{k'\in \mathcal{K}}$ in (\ref{eq:re_comp}), while $h_1(\pmb{x}_1)=2^{\frac{L_{k,n}^\cc}{B\Delta/K}}-1$ and $h_2(\pmb{x}_2)=(x^\cc_n-x^\m_k)^2+(y^\cc_n-y^\m_k)^2+H^2$ with $\pmb{x}_1=L_{k,n}^\cc$ and $\pmb{x}_2=\pmb{p}^\cc_n$ in (\ref{eq:re_comm}). Then, given a possible solution $\pmb{z}_n(v)$, we obtain a valid convex upper bound of (\ref{eq:lifetime}) by applying (\ref{eq:con_approx}) as
\begin{eqnarray}\label{eq:oma_upper}
&& \hspace{-1.3cm} E^\cc_{k,n+1}(\pmb{z}_{n+1}) + E_{\oo,k,n+2}^\cc(\pmb{z}_{n+2}) \nonumber\\
&&\hspace{-1.1cm} \le \bar{E}^\cc_{k,n+1}(\pmb{z}_{n+1};\pmb{z}_{n+1}(v)) + \bar{E}_{\oo,k,n+2}^\cc(\pmb{z}_{n+2};\pmb{z}_{n+2}(v)),
\end{eqnarray}
where $\bar{E}^\cc_{k,n}(\pmb{z}_n;\pmb{z}_n(v))$ and $\bar{E}_{\oo,k,n}^\cc(\pmb{z}_n;\pmb{z}_n(v))$ are defined in (\ref{eq:comp_approx}) and (\ref{eq:down_approx_oma}), respectively, in Appendix \ref{app:oma}, where their derivations are discussed. 

Finally, the resulting strongly convex inner approximation of (\ref{eq:CEopt}), for a given a feasible $\pmb{z}(v)=\{\pmb{z}_n(v)\}_{n \in \mathcal{N}}$, is given by
\begin{subequations}\label{eq:CEopt_Re}
\begin{eqnarray}
&& \hspace{-1.6cm} \underset{\pmb{z}}{\text{minimize}} \hspace{0.3cm} \sum_{k=1}^{K}\sum_{n=1}^{N-2} \bar{E}_{\oo,k,n}^\m(\pmb{z}_n; \pmb{z}_n(v))\label{eq:obj_Re}\\
&& \hspace{-1.5cm}  \text{s.t.} \hspace{0.2cm}  \sum_{k=1}^{K}\sum_{n=1}^{N-2}  \left(\bar{E}^\cc_{k, n+1}(\pmb{z}_{n+1}; \pmb{z}_{n+1}(v)) \right.\nonumber\\
&& \hspace{-0.7cm} \left.+ \bar{E}_{\oo,k,n+2}^\cc(\pmb{z}_{n+2};\pmb{z}_{n+2}(v)) \right.+ \sum_{n=1}^N E_{F,n}^\cc(\pmb{z}_n) \le \mathcal{E}\label{eq:lifetime_Re}\\
&& \hspace{-0.8cm}  \text{(\ref{eq:ineq_mc}) - (\ref{eq:eq_v})},
\end{eqnarray}
\end{subequations}   
which has a unique solution denoted by $\hat{\pmb{z}}(\pmb{z}(v))$. The problem (\ref{eq:CEopt_Re}) is convex. We note that closed-form solutions could be obtained via dual decomposition by following the approach in \cite{JSA17IET}, but we do not elaborate on this here given that the resulting expressions are rather cumbersome. Using (\ref{eq:CEopt_Re}), the SCA-based algorithm is summarized in Algorithm \ref{al1}. The convergence of Algorithm \ref{al1} in the sense of \cite[Theorem 2]{Scutari14Arxiv} is guaranteed if the step size sequence $\{\gamma(v)\}$ is selected such that $\gamma(v) \in (0, 1]$, $\gamma(v) \to 0$, and $\sum_{v}\gamma(v)=\infty$. More specifically, the sequence $\{\pmb{z}(v)\}$ is bounded, and every point of its limit points of $\pmb{z}(\infty)$ is a stationary solution of problem (\ref{eq:CEopt}). Furthermore, if Algorithm 1 does not stop after a finite number of steps, none of the limit points $\pmb{z}(\infty)$ is a local minimum of problem (\ref{eq:CEopt}).    

\begin{algorithm}[h]
\begin{algorithmic}
\caption{SCA-based algorithm for problem (\ref{eq:CEopt}) for orthogonal access} \label{al1} 
\State {\textbf{Input:}} $\pmb{z}(0) = \{\pmb{z}_n(0)\}_{n \in \mathcal{N}} \in \mathcal{X}$ with $\pmb{z}_n(0) \triangleq (\{L_{k,n}^\m(0)\}_{k \in \mathcal{K}}, \{l_{k,n}(0)\}_{k \in \mathcal{K}}, \{L_{k,n}^\cc(0)\}_{k \in \mathcal{K}}, \pmb{p}^\cc_n(0))$, and $\tau_{L_{k,n}^\m}, \tau_{x^\cc_n}, \tau_{y^\cc_n} > 0$ for $k \in \mathcal{K}$ and $n \in \mathcal{N}$. Set $v=0$.
\State 1. If $\pmb{z}(v)$ is a stationary solution of (\ref{eq:CEopt}), stop;
\State 2. Compute $\hat{\pmb{z}}(\pmb{z}(v))$ using (\ref{eq:CEopt_Re});
\State 3. Set $\pmb{z}(v+1)=\pmb{z}(v)+\gamma(v)(\hat{\pmb{z}}(\pmb{z}(v))-\pmb{z}(v))$ for some $\gamma(v) \in (0, 1]$;
\State 3. $v \leftarrow v+1$ and go to step 1. 
\State {\textbf{Output:}}  $\{L_{k,n}^\m\}$, $\{l_{k,n}\}$, $\{L_{k,n}^\cc\}$ and $\{\pmb{p}^\cc_n\}$.
\end{algorithmic}
\end{algorithm}

\section{Optimal Energy Consumption for Non-orthogonal Access}\label{sec:CE_noma}
In this section, we discuss the design of bit allocation and UAV trajectory for non-orthogonal access.
\subsection{Problem Formulation}\label{sec:prob_noma}
Using the same definitions as in the previous section, the problem of minimizing the total energy consumption of the MUs is formulated as in (\ref{eq:CEopt}) by substituting the energies needed for uplink and downlink communication in (\ref{eq:obj}) and (\ref{eq:lifetime}) with (\ref{eq:Ecomm_noma_up}) and (\ref{eq:Ecomm_noma_down}), respectively. We summarize the resulting problem as
\begin{subequations}\label{eq:CEopt_NOMA}
\begin{eqnarray}
&& \hspace{-1.9cm} \underset{ \{L_{k,n}^\m\}, \{l_{k,n}\}, \{L_{k,n}^\cc\}, \{\pmb{p}^\cc_n\}}{\text{minimize}} \hspace{0.3cm} \sum_{k=1}^{K}\sum_{n=1}^{N-2} E_{\nn,k,n}^\m(L_n^\m, \pmb{p}^\cc_n)\label{eq:obj_NOMA}\\
&& \hspace{-1.8cm}  \text{s.t.} \hspace{0.2cm}  \sum_{k=1}^{K}\sum_{n=1}^{N-2}  E^\cc_{k,n+1}(l_{n+1}) + E_{\nn,k,n+2}^\cc(L_{n+2}^\cc, \pmb{p}^\cc_{n+2})\nonumber\\
&& \hspace{2.1cm} + \sum_{n=1}^N E_{F,n}^\cc(\pmb{v}_n^\cc) \le \mathcal{E} \label{eq:lifetime_NOMA}\\
&& \hspace{-1.1cm}  \text{(\ref{eq:ineq_mc}) - (\ref{eq:eq_v})}.
\end{eqnarray}
\end{subequations}

\subsection{Successive Convex Approximation}\label{sec:SCA_noma} 
 
The problem (\ref{eq:CEopt_NOMA}) is non-convex due to the non-convex objective function (\ref{eq:obj_NOMA}) and the non-convex constraint (\ref{eq:lifetime_NOMA}). To address this problem, here we propose an SCA-based algorithm, for the reasons discussed in Section \ref{sec:CE_oma}. We start by rewriting the non-convex problem (\ref{eq:CEopt_NOMA}) in an equivalent non-convex form by introducing the slack variables $\alpha_{k,n} \ge 0$ and $\beta_{k,n} \ge 0$ for $n \in \mathcal{N}$ and $k \in \mathcal{K}$ as
\begin{subequations}\label{eq:opt_slack}
\begin{eqnarray}
&& \hspace{-2.1cm} \underset{\begin{subarray}{l} \{L_{k,n}^\m\}, \{l_{k,n}\}, \{L_{k,n}^\cc\},\\
\{\pmb{p}^\cc_n\}, \{\alpha_{k,n}\}, \{\beta_{k,n}\}\end{subarray}}{\text{minimize}} \hspace{0.3cm} \sum_{k=1}^{K}\sum_{n=1}^{N-2} \frac{\alpha_{k,n}}{g_{k,n}(\pmb{p}^\cc_n)}\label{eq:obj_slack}\\
&& \hspace{-2cm}  \text{s.t.} \hspace{0.2cm} \sum_{k=1}^{K}\sum_{n=1}^{N-2}  E^\cc_{k,n+1}(l_{n+1}) + \beta_{k,n+2} \nonumber\\
&& \hspace{0.3cm}+ \sum_{n=1}^N E_{F,n}^\cc(\pmb{v}_n^\cc) \le \mathcal{E} \label{eq:lifetime_slack}\\
&& \hspace{-1.3cm}  g_{k,n}(\pmb{p}^\cc_n)\hat{E}_{\nn,k,n}^\m(L_{k,n}^\m, \pmb{p}^\cc_n, \alpha_{-k, n}) \le \alpha_{k,n},\nonumber\\
&& \hspace{0.3cm} \text{for} \hspace{0.2cm} k \in \mathcal{K} \hspace{0.2cm} \text{and} \hspace{0.2cm}  n=1, \dots, N-2 \label{eq:alpha}\\
&& \hspace{-1.3cm} \hat{E}_{\nn,k,n+2}^\cc(L_{k,n+2}^\cc, \pmb{p}^\cc_{n+2}, \beta_{-k, n+2}) \le \beta_{k,n+2}, \nonumber\\
&& \hspace{0.3cm}\text{for} \hspace{0.2cm} k \in \mathcal{K} \hspace{0.2cm} \text{and} \hspace{0.2cm} n=1, \dots, N-2 \label{eq:beta}\\
&& \hspace{-1.3cm} \alpha_{k,n}, \beta_{k,n} \ge 0, \hspace{0.2cm}\text{for} \hspace{0.2cm} k \in \mathcal{K} \hspace{0.2cm} \text{and} \hspace{0.2cm} n \in \mathcal{N} \label{eq:pos_ab}\\
&& \hspace{-1.3cm}  \text{(\ref{eq:ineq_mc}) - (\ref{eq:eq_v})},
\end{eqnarray}
\end{subequations}  
where the uplink and downlink transmission energies in (\ref{eq:Ecomm_noma_up}) and (\ref{eq:Ecomm_noma_down}) are redefined with slack variables $\alpha_{-k,n}=\{\alpha_{k',n}\}_{k' \in \mathcal{K}, k' \neq k}$ and $\beta_{-k,n}=\{\beta_{k',n}\}_{k' \in \mathcal{K}, k' \neq k}$ as
\begin{subequations}
\begin{eqnarray}
&& \hspace{-0.2cm}\hat{E}_{\nn,k,n}^\m(L_{k,n}^\m, \pmb{p}^\cc_n, \alpha_{-k, n}) = \frac{1}{g_{k,n}(\pmb{p}^\cc_n)}\left(N_0 B\Delta \right. \nonumber\\
&& \hspace{0.8cm} \left.+ \sum_{k'=1, k'\neq k}^{K} \alpha_{k',n}\right)\left(2^{\frac{L_{k,n}^\m}{B\Delta}}-1\right)\\
&& \hspace{-1.3cm} \text{and}\hspace{0.5cm} \hat{E}_{\nn,k,n}^\cc(L_{k,n}^\cc, \pmb{p}^\cc_n, \beta_{-k, n}) = \left(\frac{N_0 B\Delta}{g_{k,n}(\pmb{p}^\cc_n)} \right. \nonumber\\
&& \hspace{0.8cm} \left. + \sum_{k'=1, k'\neq k}^{K} \beta_{k',n}\right)\left(2^{\frac{L_{k,n}^\cc}{B\Delta}}-1\right),
\end{eqnarray}
\end{subequations}
respectively.

In order to tackle the problem (\ref{eq:opt_slack}) via the SCA algorithm \cite{Scutari14Arxiv, Scutari16ArxivII}, as discussed in Section \ref{sec:SCA_oma}, we need to derive convex approximations for the non-convex objective function (\ref{eq:obj_slack}) and constraints (\ref{eq:lifetime_slack}), (\ref{eq:alpha}) and (\ref{eq:beta}) according to Lemma 1 and Lemma 2, respectively. To this end, let us define the set of primal variables of problem (\ref{eq:opt_slack}) as $\pmb{z}=\{\pmb{z}_n\}_{n \in \mathcal{N}}$ with $\pmb{z}_n=(\{L_{k,n}^\m\}_{k \in \mathcal{K}}, \{l_{k,n}\}_{k \in \mathcal{K}},$ $\{L_{k,n}^\cc\}_{k \in \mathcal{K}}, \pmb{p}^\cc_n, \{\alpha_{k,n}\}_{k \in \mathcal{K}}, \{\beta_{k,n}\}_{k \in \mathcal{K}})$ being the optimization variables for the $n$th frame. The objective function $\alpha_{k,n}/g_{k,n}(\pmb{p}^\cc_n)$ in (\ref{eq:obj_slack}) is the product of one non-negative linear function and one non-negative convex function, namely
\begin{subequations}
\begin{eqnarray}
&& \hspace{-1.4cm} f_1(\alpha_{k,n})= \frac{\alpha_{k,n}}{g_0},\\
&& \hspace{-2.4cm} \text{and}\hspace{0.5cm} f_2(\pmb{p}_n^\cc) =  \left(x^\cc_n-x^\m_k\right)^2+\left(y^\cc_n-y^\m_k\right)^2+H^2.
\end{eqnarray}
\end{subequations}
Therefore, using Lemma 1 and $\pmb{z}_n(v)=(\{L_{k,n}^\m(v)\}_{k \in \mathcal{K}}, \{l_{k,n}(v)\}_{k \in \mathcal{K}},\{L_{k,n}^\cc(v)\}_{k \in \mathcal{K}}, \pmb{p}^\cc_n(v),$ $\{\alpha_{k,n}(v)\}_{k \in \mathcal{K}},$ $\{\beta_{k,n}(v)\}_{k \in \mathcal{K}}) \in \mathcal{X}$ for the $v$th iterate in the feasible set $\mathcal{X}$ of (\ref{eq:opt_slack}), a strongly convex surrogate function $\bar{E}_{\nn,k,n}^\m(\pmb{z}_n; \pmb{z}_n(v))$ of the objective function $\alpha_{k,n}/g_{k,n}(\pmb{p}^\cc_n)$ in (\ref{eq:obj_slack}) is obtained as 
\begin{eqnarray}\label{eq:obj_noma_approx}
&&\hspace{-0.7cm} \bar{E}_{\nn,k,n}^\m(\pmb{z}_n; \pmb{z}_n(v)) \triangleq f_1(\alpha_{k,n})f_2(\pmb{p}^\cc_n(v))+f_1(\alpha_{k,n}(v))f_2(\pmb{p}^\cc_n)\nonumber\\
&&\hspace{0.7cm} + \frac{\tau_{\alpha_{k,n}}}{2}\left(\alpha_{k,n}-\alpha_{k,n}(v)\right)^2+\frac{\tau_{x^\cc_ n}}{2}\left(x^\cc_n-x^\cc_n(v)\right)^2\nonumber\\
&&\hspace{0.7cm} +\frac{\tau_{y^\cc_n}}{2}\left(y^\cc_n-y^\cc_n(v)\right)^2,
\end{eqnarray} 
where $\tau_{\alpha_{k,n}}, \tau_{x^\cc_n}, \tau_{y^\cc_n} > 0$, for $k \in \mathcal{K}$ and $n \in \mathcal{N}$.  

Moreover, using Lemma 2, the non-convex function $h_{k,n}^\m(L_{k,n}^\m, \alpha_{-k,n}) \triangleq g_{k,n}(\pmb{p}^\cc_n)\hat{E}_{\nn,k,n}^\m(L_{k,n}^\m, \pmb{p}^\cc_n, \alpha_{-k,n})$ in (\ref{eq:alpha}) and $\hat{E}_{\nn,k,n}^\cc(L_{k,n}^\cc, \pmb{p}^\cc_n, \beta_{-k,n})$ in the constraint (\ref{eq:beta}) can be upper bounded for a given $\pmb{z}(v)=\{\pmb{z}_n(v)\}_{n \in \mathcal{N}} \in \mathcal{X}$ as
\begin{subequations}\label{eq:noma_approx}
\begin{eqnarray}
&& \hspace{-1cm} h_{k,n}^\m(L_{k,n}^\m, \alpha_{-k,n}) \le \bar{h}_{k,n}^\m(\pmb{z}_n;\pmb{z}_n(v)) \\
&&\hspace{-2cm} \text{and}\hspace{0.5cm} \hat{E}_{\nn,k,n}^\cc(L_{k,n}^\cc, \pmb{p}^\cc_n, \beta_{-k,n}) \le \bar{E}_{\nn,k,n}^\cc(\pmb{z}_n;\pmb{z}_n(v)),
\end{eqnarray}
\end{subequations} 
where $\bar{h}_{k,n}^\m(\pmb{z}_n;\pmb{z}_n(v))$ and $\bar{E}_{\nn,k,n}^\cc(\pmb{z}_n;\pmb{z}_n(v))$ are convex functions calculated by (\ref{eq:h_noma_approx}) and (\ref{eq:up_noma_approx}), respectively, in Appendix \ref{app:noma}, where the details of the derivations are discussed. 

By using (\ref{eq:obj_noma_approx}) and (\ref{eq:noma_approx}), given a feasible $\pmb{z}(v) \in \mathcal{X}$, we have a strongly convex inner approximation of (\ref{eq:opt_slack}) as (cf. (\ref{eq:CEopt_Re}))
\begin{subequations}\label{eq:opt_slack_RE}
\begin{eqnarray}
&& \hspace{-2.6cm} \underset{\pmb{z}}{\text{minimize}} \hspace{0.3cm} \sum_{k=1}^{K}\sum_{n=1}^{N-2} \bar{E}_{\nn,k,n}^\m(\pmb{z}_n; \pmb{z}_n(v))\label{eq:obj_slack_RE}\\
&& \hspace{-2.5cm}  \text{s.t.} \hspace{0.2cm} \sum_{k=1}^{K}\sum_{n=1}^{N-2}  \bar{E}_{k, n+1}^\cc(\pmb{z}_{n+1}; \pmb{z}_{n+1}(v)) + \beta_{k,n+2} \nonumber\\
&& \hspace{-0.6cm}  + \sum_{n=1}^N E_{F,n}^\cc(\pmb{z}_n)  \le \mathcal{E} \label{eq:lifetime_slack_RE}\\
&& \hspace{-1.8cm}  \bar{h}_{k,n}^\m(\pmb{z}_n; \pmb{z}_n(v)) \le \alpha_{k,n}, \nonumber\\
&& \hspace{-0.6cm} \text{for} \hspace{0.2cm} k \in \mathcal{K} \hspace{0.2cm} \text{and} \hspace{0.2cm} n=1, \dots, N-2 \label{eq:alpha_RE}\\
&& \hspace{-1.8cm} \bar{E}_{\nn,k,n+2}^\cc(\pmb{z}_{n+2}; \pmb{z}_{n+2}(v)) \le \beta_{k,n+2}, \nonumber\\
&& \hspace{-0.6cm}\text{for} \hspace{0.2cm} k \in \mathcal{K} \hspace{0.2cm} \text{and} \hspace{0.2cm} n=1, \dots, N-2\label{eq:beta_RE}\\
&& \hspace{-1.8cm}  \text{(\ref{eq:pos_ab}), (\ref{eq:ineq_mc}) - (\ref{eq:eq_v})},
\end{eqnarray}
\end{subequations}  
where $\bar{E}_{k,n}^\cc(\pmb{z}_n; \pmb{z}_n(v))$ is defined equivalently in (\ref{eq:comp_approx}), which provides a unique solution denoted by $\hat{\pmb{z}}(\pmb{z}(v))$. The SCA-based algorithm is summarized using (\ref{eq:opt_slack_RE}) in Algorithm \ref{al2}. Its convergence is established by following \cite[Theorem 2]{Scutari14Arxiv} as discussed in Section \ref{sec:CE_oma}.    

\begin{algorithm}[h]
\begin{algorithmic}
\caption{SCA-based algorithm for problem (\ref{eq:opt_slack}) for non-orthogonal access} \label{al2} 
\State {\textbf{Input:}} $\pmb{z}(0) = \{\pmb{z}_n(0)\}_{n \in \mathcal{N}} \in \mathcal{X}$ with $\pmb{z}_n(0) \triangleq (\{L_{k,n}^\m(0)\}_{k \in \mathcal{K}}, \{l_{k,n}(0)\}_{k \in \mathcal{K}}, \{L_{k,n}^\cc(0)\}_{k \in \mathcal{K}}, \pmb{p}^\cc_n(0),$ $\{\alpha_{k,n}(0)\}_{k \in \mathcal{K}}, \{\beta_{k,n}(0)\}_{k \in \mathcal{K}})$, and $\tau_{\alpha_{k,n}}, \tau_{x^\cc_n}, \tau_{y^\cc_n} > 0$ for $k \in \mathcal{K}$ and $n \in \mathcal{N}$. Set $v=0$.
\State 1. If $\pmb{z}(v)$ is a stationary solution of (\ref{eq:opt_slack}), stop;
\State 2. Compute $\hat{\pmb{z}}(\pmb{z}(v))$ using (\ref{eq:opt_slack_RE});
\State 3. Set $\pmb{z}(v+1)=\pmb{z}(v)+\gamma(v)(\hat{\pmb{z}}(\pmb{z}(v))-\pmb{z}(v))$ for some $\gamma(v) \in (0, 1]$;
\State 3. $v \leftarrow v+1$ and go to step 1. 
\State {\textbf{Output:}}  $\{L_{k,n}^\m\}$, $\{l_{k,n}\}$, $\{L_{k,n}^\cc\}$, $\{\pmb{p}^\cc_n\}$, $\{\alpha_{k,n}\}$ and $\{\beta_{k,n}\}$.
\end{algorithmic}
\end{algorithm}

\section{UAV's Propulsion Energy Consumption}\label{sec:pro}
In the previous sections, we assumed the UAV's energy consumption model (\ref{eq:pro_simple}) for flying, in which the flying energy depends only on the velocity. In this section, we adopt a more refined model following \cite{Zeng16Arxiv2, Filippone06, Leishman06, Kong09AIAA}, in which the propulsion energy of the UAV depends on both the velocity and acceleration vectors. One of the goals of this study is to understand the impact of the energy consumption model on the optimal system design.      

Let us denote the UAV's acceleration vector for the $n$th frame as $\pmb{a}_n^\cc$, where
\begin{equation}\label{eq:a}
\pmb{a}_{n}^\cc = \frac{\pmb{v}_{n+1}^\cc-\pmb{v}_{n}^\cc}{\Delta}.
\end{equation}
Following \cite{Zeng16Arxiv2, Filippone06, Leishman06, Kong09AIAA}, the UAV's propulsion energy consumption at the $n$th frame can be modeled as 
\begin{eqnarray}\label{eq:pro}
&&\hspace{-1cm}\text{(Model 2)}\nonumber\\
&&\hspace{-1cm} E_{F, n}^\cc(\pmb{v}_n^\cc, \pmb{a}_n^\cc) = \kappa_1 \left\|\pmb{v}_n^\cc\right\|^3 + \frac{\kappa_2}{\left\|\pmb{v}_n^\cc\right\|}\left(1+\frac{\left\|\pmb{a}_n^\cc\right\|^2}{g^2}\right),
\end{eqnarray} 
 where $g$ is gravitational acceleration. A discussion of model (\ref{eq:pro}) can be found along with the values for the constants $\kappa_1$ and $\kappa_2$ in in Appendix \ref{app:Ef}. The velocity vector $\pmb{v}_n^\cc$ and acceleration vector $\pmb{a}_n^\cc$ are related to the UAV's position $\pmb{p}_n^\cc$ according to the second-order Taylor approximation model
\begin{equation}\label{eq:rel_state}
\pmb{p}_{n+1}^\cc = \pmb{p}_{n}^\cc + \pmb{v}_{n}^\cc\Delta+\frac{1}{2}\pmb{a}_{n}^\cc\Delta^2,
\end{equation} 
for $n \in \mathcal{N}$.

Considering an overall constraint on the UAV energy with (\ref{eq:pro}) in lieu of (\ref{eq:pro_simple}) yields the following optimization problem for orthogonal access
\begin{subequations}\label{eq:CEopt_pro}
\begin{eqnarray}
&& \hspace{-1.5cm} \underset{\substack{ \{L_{k,n}^\m\}, \{l_{k,n}\}, \{L_{k,n}^\cc\}, \\ \{\pmb{p}_n^\cc\}, \{\pmb{v}_n^\cc\}, \{\pmb{a}_n^\cc\}} }{\text{minimize}} \hspace{0.3cm} \sum_{k=1}^{K}\sum_{n=1}^{N-2} E_{\oo,k,n}^\m(L_{k,n}^\m,\pmb{p}^\cc_n)\label{eq:obj_pro}\\
&& \hspace{-1.4cm}  \text{s.t.} \hspace{0.2cm} \sum_{k=1}^{K}\sum_{n=1}^{N-2}  E^\cc_{k,n+1}(l_{n+1})  + E_{\oo,k,n+2}^\cc(L_{k,n+2}^\cc, \pmb{p}^\cc_{n+2})\nonumber\\
&& \hspace{2.5cm} +  \sum_{n=1}^{N}  E^\cc_{F,n}(\pmb{v}_n^\cc, \pmb{a}_n^\cc) \le \mathcal{E}\label{eq:lifetime_pro}\\
&& \hspace{-0.7cm} \pmb{v}_{n+1}^\cc = \pmb{v}_{n}^\cc + \pmb{a}_{n}^\cc\Delta, \hspace{0.2cm} \text{for} \hspace{0.2cm} n \in \mathcal{N}\label{eq:rel_state_v}\\
&& \hspace{-0.7cm} \pmb{p}_{n+1}^\cc = \pmb{p}_{n}^\cc + \pmb{v}_{n}^\cc\Delta+\frac{1}{2}\pmb{a}_{n}^\cc\Delta^2, \hspace{0.2cm} \text{for} \hspace{0.2cm} n \in \mathcal{N}\label{eq:rel_state_p}\\
&& \hspace{-0.7cm}  \pmb{v}_1^\cc = \pmb{v}_{N+1}^\cc = \pmb{v}^\cc \label{eq:initv}\\
&& \hspace{-0.75cm} \left\|\pmb{a}_n^\cc\right\| \le a_{\max}, \hspace{0.2cm} \text{for} \hspace{0.2cm} n \in \mathcal{N} \label{eq:ineq_a}\\
&& \hspace{-0.7cm} \text{(\ref{eq:ineq_mc}) - (\ref{eq:ineq_v})},  \label{eq:whole_oma}
\end{eqnarray}
\end{subequations}    
where (\ref{eq:lifetime_pro}) is the overall UAV energy constraint; (\ref{eq:initv}) represents the UAV's initial and final velocity constraint; and (\ref{eq:ineq_a}) guarantees a maximum acceleration constraint of $a_{\max}$. Note that, as compared to (\ref{eq:CEopt}), problem (\ref{eq:CEopt_pro}) has the additional optimization variables $\{\pmb{v}_n^\cc\}$ and $\{\pmb{a}_n^\cc\}$. 

To tackle the non-convex problem (\ref{eq:CEopt_pro}), we apply the SCA approach as above in Section \ref{sec:SCA_oma}. The key difference with respect to Section \ref{sec:SCA_oma} is the need to cope with the non-convex function $E^\cc_{F,n}(\pmb{v}_n^\cc, \pmb{a}_n^\cc)$ in (\ref{eq:lifetime_pro}). To elaborate, we introduce nonnegative slack variables $\{\tau_{v_n^\cc} \ge 0\}$, and impose the additional constraints $\|\pmb{v}_n^\cc\| \ge \tau_{v_n^\cc}$ for $n \in \mathcal{N}$. Under these constraints, the propulsion energy consumption $E^\cc_{F,n}(\pmb{v}_n^\cc, \pmb{a}_n^\cc)$ in (\ref{eq:pro}) is upper bounded as
\begin{eqnarray}\label{eq:pro_re}
\hspace{-2.3cm} E^\cc_{F,n}(\pmb{v}_n^\cc, \pmb{a}_n^\cc) &\le& \kappa_1 \left\|\pmb{v}_n^\cc\right\|^3 + \frac{\kappa_2}{\tau_{v_n^\cc}}+\frac{\kappa_2\left\|\pmb{a}_n^\cc\right\|^2}{\tau_{v_n^\cc} g^2} \nonumber\\
&\triangleq& \bar{E}_{F, n}^\cc(\pmb{z}_n; \pmb{z}_n(v)),
\end{eqnarray} 
where the inequality in (\ref{eq:pro_re}) results from the constraint $\|\pmb{v}_n^\cc\|^2 \ge \tau_{v_n^\cc}^2$, yielding the convex upper bound $\bar{E}_{F, n}^\cc(\pmb{z}_n; \pmb{z}_n(v))$. In (\ref{eq:pro_re}), we redefined the set of variables $\pmb{z}$ and $\pmb{z}_n(v)$ by including the additional variables $\{\pmb{v}_n^\cc\}$, $\{\pmb{a}_n^\cc\}$ and $\{\tau_{v_n^\cc}\}$ as $\pmb{z}=\{\pmb{z}_n\}_{n \in \mathcal{N}}$ with $\pmb{z}_n=(\{L_{k,n}^\m\}_{k \in \mathcal{K}}, \{l_{k,n}\}_{k \in \mathcal{K}}, \{L_{k,n}^\cc\}_{k \in \mathcal{K}}, \pmb{p}_n^\cc, \pmb{v}_n^\cc,$ $\pmb{a}_n^\cc, \tau_{v_n^\cc})$ and as $\pmb{z}_n(v)=(\{L_{k,n}^\m(v)\}_{k \in \mathcal{K}}, \{l_{k,n}(v)\}_{k \in \mathcal{K}}, \{L_{k,n}^\cc(v)\}_{k \in \mathcal{K}},$ $\pmb{p}_n^\cc(v), \pmb{v}_n^\cc(v), \pmb{a}_n^\cc(v), \tau_{v_n^\cc}(v)) \in \mathcal{X}$ for the $v$th iterate, where $\mathcal{X}$ is the feasible set of problem (\ref{eq:CEopt_pro}). By using the bound (\ref{eq:pro_re}), we obtain the convex program to be solved at the $v$th iteration as
\begin{subequations}\label{eq:CEopt_Re_pro}
\begin{eqnarray}
&& \hspace{-1.2cm} \underset{\pmb{z}}{\text{minimize}} \hspace{0.3cm} \sum_{k=1}^{K}\sum_{n=1}^{N-2} \bar{E}_{\oo,k,n}^\m(\pmb{z}_n; \pmb{z}_n(v))\label{eq:obj_Re_pro}\\
&& \hspace{-1.1cm}  \text{s.t.} \hspace{0.2cm}  \sum_{k=1}^{K}\sum_{n=1}^{N-2}  \left(\bar{E}^\cc_{k, n+1}(\pmb{z}_{n+1}; \pmb{z}_{n+1}(v)) \right.\nonumber\\ 
&& \hspace{-1.2cm} \left. + \bar{E}_{\oo,k,n+2}^\cc(\pmb{z}_{n+2};\pmb{z}_{n+2}(v))\right) + \sum_{n=1}^{N}  \bar{E}^\cc_{F,n}(\pmb{z}_n; \pmb{z}_n(v)) \le \mathcal{E} \nonumber\\
&&  \label{eq:lifetime_Re_pro}\\
&& \hspace{-0.5cm}  \tau_{v_n^\cc}^2 \le f^{LB}(\pmb{z}_n; \pmb{z}_n(v)), \hspace{0.2cm} \text{for} \hspace{0.2cm} n \in \mathcal{N} \label{eq:LB}\\
&& \hspace{-0.5cm}  \tau_{v_n^\cc} \ge 0, \hspace{0.2cm} \text{for} \hspace{0.2cm} n \in \mathcal{N} \label{eq:pos_tau}\\
&& \hspace{-0.5cm}  \text{(\ref{eq:rel_state_v}) - (\ref{eq:whole_oma})}, \label{eq:whole_noma}
\end{eqnarray}
\end{subequations}
where $f^{LB}(\pmb{z}_n; \pmb{z}_n(v))$ is the linear lower bound on the squared norm $\|\pmb{v}_n^\cc\|^2$ as
\begin{eqnarray}
f^{LB}(\pmb{z}_n; \pmb{z}_n(v)) &=& \left\|\pmb{v}_n^\cc(v)\right\|^2 + 2\left(\pmb{v}_n^\cc(v)\right)^T\left(\pmb{v}_n^\cc-\pmb{v}_n^\cc(v)\right) \nonumber\\
&\le& \|\pmb{v}_n^\cc\|^2.
\end{eqnarray}
The problem (\ref{eq:CEopt_Re_pro}) is used within Algorithm \ref{al1}, where (\ref{eq:CEopt}) and (\ref{eq:CEopt_Re}) is substituted with (\ref{eq:CEopt_pro}) and (\ref{eq:CEopt_Re_pro}), respectively, to yield the proposed SCA solution. 

In a similar manner, we can consider non-orthogonal access yielding the problem  
\begin{subequations}\label{eq:CEopt_NOMA_pro}
\begin{eqnarray}
&& \hspace{-1.9cm} \underset{\substack{ \{L_{k,n}^\m\}, \{l_{k,n}\}, \{L_{k,n}^\cc\}, \\ \{\pmb{p}_n^\cc\}, \{\pmb{v}_n^\cc\}, \{\pmb{a}_n^\cc\}} }{\text{minimize}} \hspace{0.3cm} \sum_{k=1}^{K}\sum_{n=1}^{N-2} E_{\nn,k,n}^\m(L_n^\m, \pmb{p}^\cc_n)\label{eq:obj_NOMA_pro}\\
&& \hspace{-1.8cm}  \text{s.t.} \hspace{0.2cm} \sum_{k=1}^{K}\sum_{n=1}^{N-2}  E^\cc_{k,n+1}(l_{n+1}) + E_{\nn,k,n+2}^\cc(L_{n+2}^\cc, \pmb{p}^\cc_{n+2}) \nonumber\\
&& \hspace{2.1cm} + \sum_{n=1}^{N}  E^\cc_{F,n}(\pmb{v}_n^\cc, \pmb{a}_n^\cc) \le \mathcal{E} \label{eq:lifetime_NOMA_pro}\\
&& \hspace{-1.2cm}  \text{(\ref{eq:rel_state_v}) - (\ref{eq:whole_oma})},
\end{eqnarray}
\end{subequations}
where (\ref{eq:lifetime_NOMA_pro}) is the overall UAV energy constraint. Then, using slack variables $\alpha_{k,n} \ge 0$ and $\beta_{k,n} \ge 0$ for $k \in \mathcal{K}$ and $n \in \mathcal{N}$ as in (\ref{eq:opt_slack}), we can rewrite the problem (\ref{eq:CEopt_NOMA_pro}) into 
\begin{subequations}\label{eq:opt_slack_pro}
\begin{eqnarray}
&& \hspace{-2.1cm} \underset{\begin{subarray}{l} \{L_{k,n}^\m\}, \{l_{k,n}\}, \{L_{k,n}^\cc\},\\
\{\pmb{p}^\cc_n\}, \{\pmb{v}_n^\cc\}, \{\pmb{a}_n^\cc\}, \\ \{\alpha_{k,n}\}, \{\beta_{k,n} \}\end{subarray}}{\text{minimize}} \hspace{0.3cm} \sum_{k=1}^{K}\sum_{n=1}^{N-2} \frac{\alpha_{k,n}}{g_{k,n}(\pmb{p}^\cc_n)}\label{eq:obj_slack_pro}\\
&& \hspace{-2cm}  \text{s.t.} \hspace{0.2cm} \sum_{k=1}^{K}\sum_{n=1}^{N-2}  E^\cc_{k,n+1}(l_{n+1}) + \beta_{k,n+2} \nonumber\\
&& \hspace{0.3cm}+ \sum_{n=1}^{N}  E^\cc_{F,n}(\pmb{v}_n^\cc, \pmb{a}_n^\cc) \le \mathcal{E} \label{eq:lifetime_slack_pro}\\
&& \hspace{-1.3cm}  \text{(\ref{eq:rel_state_v}) - (\ref{eq:whole_oma}), (\ref{eq:alpha}) - (\ref{eq:pos_ab})}.
\end{eqnarray}
\end{subequations} 
This can be tackled using SCA in Algorithm \ref{al2} with the following convex problem as
\begin{subequations}\label{eq:opt_slack_RE_pro}
\begin{eqnarray}
&& \hspace{-2.6cm} \underset{\pmb{z}}{\text{minimize}} \hspace{0.3cm} \sum_{k=1}^{K}\sum_{n=1}^{N-2} \bar{E}_{\nn,k,n}^\m(\pmb{z}_n; \pmb{z}_n(v))\label{eq:obj_slack_RE_pro}\\
&& \hspace{-2.5cm}  \text{s.t.} \hspace{0.2cm} \sum_{k=1}^{K}\sum_{n=1}^{N-2} \bar{E}_{k, n+1}^\cc(\pmb{z}_{n+1}; \pmb{z}_{n+1}(v)) + \beta_{k,n+2} \nonumber\\
&& \hspace{0.5cm} + \sum_{n=1}^{N} \bar{E}^\cc_{F,n}(\pmb{z}_n; \pmb{z}_n(v)) \le \mathcal{E} \label{eq:lifetime_slack_RE_pro}\\
&& \hspace{-1.9cm}  \text{(\ref{eq:LB}) - (\ref{eq:whole_noma}), (\ref{eq:alpha_RE}), (\ref{eq:beta_RE}), (\ref{eq:pos_ab})},
\end{eqnarray}
\end{subequations}    
in lieu of (\ref{eq:opt_slack}) and (\ref{eq:opt_slack_RE}), respectively, where $\pmb{z}=\{\pmb{z}_n\}_{n \in \mathcal{N}}$ with $\pmb{z}_n=(\{L_{k,n}^\m\}_{k \in \mathcal{K}}, \{l_{k,n}\}_{k \in \mathcal{K}},$ $\{L_{k,n}^\cc\}_{k \in \mathcal{K}},\pmb{p}_n^\cc,\pmb{v}_n^\cc, \pmb{a}_n^\cc, \{\alpha_{k,n}\}_{k \in \mathcal{K}},\{\beta_{k,n}\}_{k \in \mathcal{K}}, \tau_{v_n^\cc})$; $\pmb{z}_n(v)=(\{L_{k,n}^\m(v)\}_{k \in \mathcal{K}},\{l_{k,n}(v)\}_{k \in \mathcal{K}},\{L_{k,n}^\cc(v)\}_{k \in \mathcal{K}},\pmb{p}_n^\cc(v), \pmb{v}_n^\cc(v),$ $\pmb{a}_n^\cc(v), \{\alpha_{k,n}(v)\}_{k \in \mathcal{K}},\{\beta_{k,n}(v)\}_{k \in \mathcal{K}},$ $\tau_{v_n^\cc}(v))$ $\in \mathcal{X}$ with the feasible set $\mathcal{X}$; and $E^\cc_{F,n}(\pmb{v}_n^\cc, \pmb{a}_n^\cc)$ and $\bar{E}^\cc_{F,n}(\pmb{z}_n; \pmb{z}_n(v))$ are defined in (\ref{eq:pro}) and (\ref{eq:pro_re}), respectively.

\section{Numerical Results}\label{sec:num}
In this section, we evaluate the performance of the proposed optimization algorithm over bit allocation and UAV's trajectory via numerical experiments. We will consider both the results of the optimization studied in Section \ref{sec:CE_oma} and Section \ref{sec:CE_noma} in which the UAV energy for flying is given by (\ref{eq:pro_simple}) (Model 1) or (\ref{eq:pro}) (Model 2). Furthermore, for reference, we consider the following schemes: ($i$) \textit{No optimization}: With this scheme, the same number of bits is transmitted in uplink and downlink in each frame, the same number of bits is computed at the cloudlet at each frame, and the cloudlet flies at constant velocity between the initial and final positions, i.e., $L_{k,n}^\m=l_{k,n+1}=I_k/(N-2)$ and $L_{k,n+2}^\cc=I_kO_k/(N-2)$ for $k \in \mathcal{K}$ and $n = 1, \dots, N-2$, and $x_n^\cc=x_I^\cc+(n-1)(x_{F}^\cc-x_I^\cc)/N$ and $y_n^\cc=y_I^\cc+(n-1)(y_F^\cc-y_I^\cc)/N$ for $n \in \mathcal{N}$; ($ii$) \textit{Optimized bit allocation}: With this scheme, the optimized number of bits is transmitted in each uplink and downlink frame and computed at the cloudlet by the proposed algorithms while keeping the described constant-velocity cloudlet's trajectory; ($iii$) \textit{Optimized UAV's trajectory}: With this scheme, the cloudlet flies along the optimized trajectory between the initial and final positions as obtained by the proposed algorithms with fixed equal bit allocation in each frame. The UAV's initial and final velocity constraint for Model 2 is set to be $\pmb{v}^\cc=\|\pmb{v}^\cc\|(\pmb{p}_F^\cc-\pmb{p}_I^\cc)/\|\pmb{p}_F^\cc-\pmb{p}_I^\cc\|$, where $\|\pmb{v}^\cc\| \le v_{\max}$ is its initial and final speed. The remaining parameters used in the simulations, unless specified otherwise, are summarized in Table \ref{t1}, where $\kappa_1$ and $\kappa_2$ are set for Model 2 by considering the fixed-wing UAV's parameters.     

\begin{table}[t]
\caption{Simulation Parameters} \label{t2}
\begin{center}
    \begin{tabular}{|p{1cm}|p{2.5cm}|p{1cm}|p{2.5cm}|}
    \hline
    Parameter & Value & Parameter & Value\\ \hline\hline
    $B$ & $40$ MHz & $N_0$ & $-174$ dBm/Hz \\ \hline
    $\gamma_k^\m$, $\gamma^\cc$ & $10^{-28}$ \cite{Yuan03ACM, Yuan06ACM} & $O_k$ & $0.5$\\ \hline
    $C_k$ & $1550.7$ ($95$th percentile of random $C_k$ in \cite{Yuan03ACM, Yuan06ACM}) & $H$ & $5$ m \\ \hline
    $\mathcal{E}$ & $500$ kJ & $g$ & $9.8$ m/s$^2$ \\ \hline
    $v_{\max}$ & $50$ m/s & $a_{\max}$ & $30$ m/s$^2$\\ \hline
    $\Delta$ & $45$ ms & $M$ & $9.65$ kg \\ \hline
	$\rho$ & $1.225$ kg/m$^3$ & $C_{D_0}$ & $0.0355$ \\ \hline    
    $S_r$ & $3.77$ m$^2$ & $e_0$ & $0.85$ \\ \hline
    $A_R$ & $13$  & $\kappa$ & $0.2171$ \\ \hline
    $\kappa_1$ & $0.0037$ & $\kappa_2$ & $5.0206$\\
\hline
    \end{tabular}
    \end{center}
\end{table} 

\begin{figure}[t]
\begin{center}
\includegraphics[width=9cm]{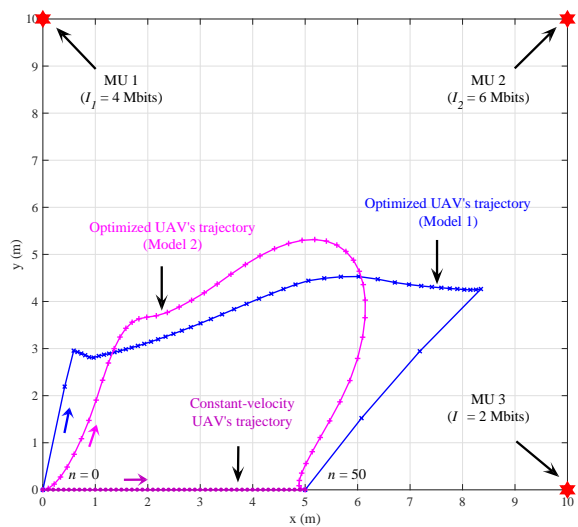}
\caption{Position of the MUs and optimized UAV's trajectory for orthogonal access with Algorithm \ref{al1} ($K=3$, $T=2.25$ s, $(I_1, I_2, I_3)=(4,6,2)$ Mbits, $\pmb{p}_1^\m=(0,10,0)$ m, $\pmb{p}_2^\m=(10,10,0)$ m, $\pmb{p}_3^\m=(10,0,0)$ m, $\pmb{p}_I^\cc=(0,0)$ m, $\pmb{p}_F^\cc=(5,0)$ m and the reference SNR $g_0/(N_0B)=-5$ dB).} \label{fig:oma_xy}
\end{center}
\end{figure}

\begin{figure}[t]
\begin{center}
\includegraphics[width=9cm]{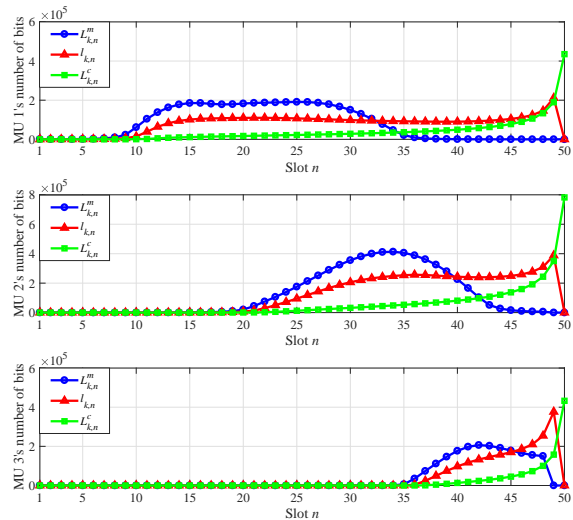}
\caption{Optimized bit allocation for the UAV trajectory under Model 2 in Fig. \ref{fig:oma_xy} ($K=3$, $T=2.25$ s, $(I_1, I_2, I_3)=(4,6,2)$ Mbits, $\pmb{p}_1^\m=(0,10,0)$ m, $\pmb{p}_2^\m=(10,10,0)$ m, $\pmb{p}_3^\m=(10,0,0)$ m, $\pmb{p}_I^\cc=(0,0)$ m, $\pmb{p}_F^\cc=(5,0)$ m and the reference SNR $g_0/(N_0B)=-5$ dB).} \label{fig:oma_bits}
\end{center}
\end{figure}

As shown in Fig. \ref{fig:oma_xy}, in the first scenario under study, there are $K=3$ MUs located at positions $\pmb{p}_1^\m=(0,10,0)$, $\pmb{p}_2^\m=(10,10,0)$ and $\pmb{p}_3^\m=(10,0,0)$, while the initial and final positions of the cloudlet are $\pmb{p}_I^\cc=(0,0)$ to $\pmb{p}_F^\cc=(5,0)$, respectively, with the UAV's initial speed $\|\pmb{v}^\cc\|=2.22$ m/s. The numbers of bits to be offloaded in the uplink from the MUs are assumed to be $I_1=4$ Mbits, $I_2=6$ Mbits and $I_3=2$ Mbits. The latency constraint is $T=2.25$ s, or $N=50$ with the parameters in Table \ref{t1}, and the reference SNR $g_0/(N_0B)=-5$ dB.

Fig. \ref{fig:oma_xy} shows the optimized trajectories obtained for orthogonal access under both UAV's flying energy consumption models. The same qualitative behavior was observed for non-orthogonal access with Algorithm \ref{al2} (not reported here). Fig. \ref{fig:oma_xy} shows that, under both models, the UAV tends to stay longer near MU $2$, which has the largest number of input bits to offload. However, when including the UAV's propulsion energy consumption as in Model 2, the trajectory tends to turn smoothly compared to Model 1 in order to limit the energy consumption caused by accelerations. This demonstrates the impact of the energy consumption model on the optimal system design.    

For the same example, Fig. \ref{fig:oma_bits} shows the optimized bit allocation for the UAV trajectory in Fig. \ref{fig:oma_xy} that is attained under Model 2. A similar trend is observed also under Model 1 (not shown here). It is seen that, when the UAV is closer to an MU $k$, a larger number $\{L_{k,n}^\m\}$ of bits for uplink transmission is allocated for MU $k$. Moreover, the bit allocation $\{l_{k,n}\}$ for computation and $\{L_{k,n}^\cc\}$ for downlink transmission are constrained by the number of bits received in the uplink and on the output bits obtained as a result of computing, respectively. Finally, the downlink bit allocation $\{L_{k,n}^\cc\}$ is seen to be less affected by the cloudlet's position compared to the uplink bit allocation $\{L_{k,n}^\m\}$ since the algorithm does not attempt to minimize UAV's energy consumption but it only imposes the UAV energy budget at the cloudlet. 

\begin{figure}[t]
\begin{center}
\includegraphics[width=9cm]{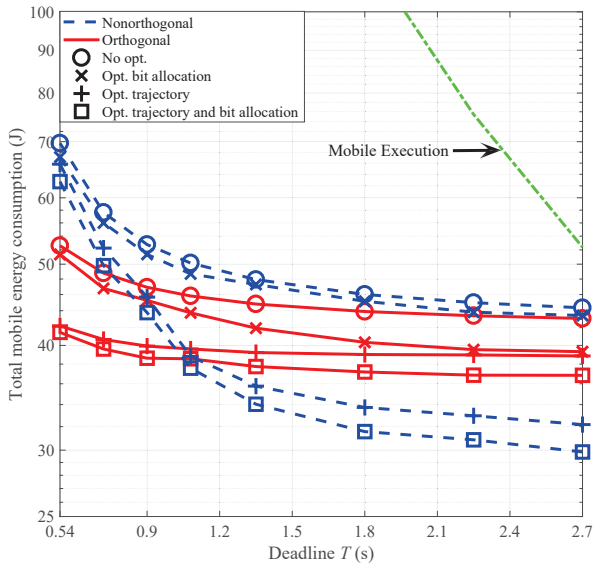}
\caption{Average total energy consumption of the MUs as a function of the deadline $T$ under Model 1 when the MUs are uniformly distributed in a $10 \times 10$ m$^2$ square region ($K=2$, $(I_1, I_2)=(8,8)$ Mbits, $\pmb{p}_I^\cc=(0,0)$ m, $\pmb{p}_F^\cc=(0,8)$ m and the reference SNR $g_0/(N_0B)=-2.5$ dB).} \label{fig:T}
\end{center}
\end{figure}

Fig. \ref{fig:T} compares the average total energy consumptions (\ref{eq:optE_ME_tot}) for mobile execution with the mobile energy needed for offloading using orthogonal and non-orthogonal access as a function of the deadline $T$ under Model 1. For this experiment, we have $K=2$ MUs with input bits $I_1=I_2=8$ Mbits that are uniformly distributed in a $10 \times 10$ m$^2$ square region. We assume the initial and final position of cloudlet as $\pmb{p}_I^\cc=(0,0)$ and $\pmb{p}_F^\cc=(0,8)$, respectively. The energy shown in Fig. \ref{fig:T} is averaged with respect to the MUs' locations. The reference SNR $g_0/(N_0B)$ is set to be $-2.5$ dB. From Fig. \ref{fig:T}, we first observe that as the deadline $T$ becomes more stringent, the energy savings of cloudlet offloading become more prominent compared with respect to mobile execution given that mobile computing energy grows as $T^{-2}$ as per (\ref{eq:optE_ME_tot}) while the mobile energy with offloading decreases more slowly with $T$. Furthermore, we note the significant gains obtained by means of joint optimization of trajectory and bit allocation. For instance, for $T=2.7$ s, the proposed scheme requires an average total MUs' energy consumption of $36.8$ J for orthogonal access and $29.9$ J for non-orthogonal access, whereas the non-optimized systems with equal bit allocation and constant-velocity cloudlet trajectory requires $43.1$ J and $44.3$ J, respectively, which implies a $14.5$\% and $32.7$\% decrease on the mobile energy consumption. The larger gain for non-orthogonal access can be attributed to the dependence of its performance on the mutual interference among MUs, which is affected by bit allocation. Also, optimizing the trajectory is seen to be more advantageous than optimizing only the bit allocation. For instance, of the mentioned $32.7$\% decrease in energy with non-orthogonal access, $27.4$\% can be obtained by optimizing only the trajectory, while $2$\% is achieved by optimizing only the bit allocation. Finally, upon optimization, non-orthogonal access is preferred to the orthogonal access unless $T$ is small. This can be explained since a shorter deadline $T$ requires a larger energy consumption, which renders the performance of non-orthogonal access interference-limited. Note that if the deadline $T$ is not enough long for the UAV to fly from its initial to final location under its maximum velocity constraint, the offloading becomes infeasible (cf. (\ref{eq:vmax_tot})).    

\section{Concluding Remarks}\label{sec:con}
In this paper, we studied a mobile cloud computing architecture based on a UAV-mounted cloudlet which provides the offloading opportunities to multiple static mobile devices. Two types of access schemes, namely orthogonal access and non-orthogonal access, were considered for the uplink and downlink transmissions required for the offloading procedure. We tackled the minimization of the mobile energy over the bit allocation for uplink, downlink and computation as well as over the UAV's trajectory for both access schemes by means of successive convex approximation methods. Numerical results verify the significant mobile energy savings of the proposed joint optimization of bit allocation and cloudlet's trajectory as compared to local mobile execution, as well as to partial optimization approaches that design only the bit allocation or the cloudlet's trajectory. They also point to the importance of acquiring accurate energy consumption models for the UAV. Interesting open problems concern the generalization of the optimization studied here to multiple moving interfering mobile devices and to trajectories with a variable altitude.
\appendices
\section{Derivations of (\ref{eq:oma_upper})}\label{app:oma}
In this appendix, for a given $\pmb{z}(v) \in \mathcal{X}$ with the feasible set $\mathcal{X}$ of problem (\ref{eq:CEopt}), we derive the convex upper bounds $\bar{E}^\cc_{k,n}(\pmb{z}_n;\pmb{z}_n(v))$ and $\bar{E}_{\oo,k,n}^\cc(\pmb{z}_n;\pmb{z}_n(v))$ of non-convex functions $E^\cc_{k,n}(\pmb{z}_n)$ and $E_{\oo,k,n}^\cc(\pmb{z}_n)$, respectively, in (\ref{eq:lifetime}) by following Lemma 2.

The computing energy consumption $E^\cc_{k,n}(\pmb{z}_{n})$ of MU $k$ can be first rewritten as 
\begin{eqnarray}
&&\hspace{-2cm}E^\cc_{k,n}(\pmb{z}_n) =\frac{\gamma^\cc C_k}{\Delta^2}\left[\frac{1}{2}\left(l_{k,n}+\left(\sum_{k'=1}^{K}C_{k'}l_{k',n}\right)^2\right)^2 \right.\nonumber\\
&& \left. -\frac{1}{2}\left(\left(l_{k,n}\right)^2+\left(\sum_{k'=1}^{K}C_{k'}l_{k',n}\right)^4\right)\right], 
\end{eqnarray} 
which leads to the convex upper bound of $E^\cc_{k,n}(\pmb{z}_n)$ around $\pmb{z}_{n}(v)$ as    
\begin{eqnarray}\label{eq:comp_approx}
&& \hspace{-1.1cm} \bar{E}^\cc_{k,n}(\pmb{z}_n; \pmb{z}_n(v)) \triangleq E^\cc_{k,n}(l_{n}; l_{n}(v)) \nonumber\\
&&\hspace{-1.1cm} = \frac{\gamma^\cc C_k}{2\Delta^2}\left[\left(l_{k,n}+\left(\sum_{k'=1}^{K}C_{k'}l_{k',n}\right)^2\right)^2 \right.\nonumber\\
&&\hspace{-0.8cm} \left. -\left(l_{k,n}(v)\right)^2-\left(\sum_{k'=1}^{K}C_{k'}l_{k',n}(v)\right)^4\right] \nonumber\\
&&\hspace{-1.1cm} - \frac{\gamma^\cc C_k}{\Delta^2}\left[l_{k,n}(v)\left(l_{k,n}-l_{k,n}(v)\right)+2\left(\sum_{k'=1}^{K}C_{k'}l_{k',n}(v)\right)^3 \right. \nonumber\\
&&\hspace{-0.8cm} \left. \times\left(\sum_{k'=1}^{K}C_{k'}\left(l_{k',n}-l_{k',n}(v)\right)\right)\right].
\end{eqnarray}
Similarly, we can rewrite the downlink communication energy consumption $E_{\oo,k,n}^\cc(\pmb{z}_n)$ as
\begin{eqnarray}
&& \hspace{-1.5cm} E_{\oo,k,n}^\cc(\pmb{z}_n)= \frac{N_0 B \Delta/K}{g_0}\left[\frac{1}{2}\left(2^{\frac{L_{k,n}^\cc}{B\Delta/K}}-1 \right.\right.\nonumber\\
&& \hspace{0.4cm} \left.\left. +\left(x^\cc_n-x^\m_k\right)^2+\left(y^\cc_n-y^\m_k\right)^2+H^2\right)^2\right.\nonumber\\
&& \hspace{0.1cm} \left.-\frac{1}{2}\left(\left(2^{\frac{L_{k,n}^\cc}{B\Delta/K}}-1\right)^2 \right.\right.\nonumber\\
&& \hspace{0.4cm} \left.\left. +\left(\left(x^\cc_n-x^\m_k\right)^2+\left(y^\cc_n-y^\m_k\right)^2+H^2\right)^2\right)\right].
\end{eqnarray}
Then, the desired convex upper bound of $E_{\oo,k,n}^\cc(\pmb{z}_n)$ around $\pmb{z}_n(v)$ can then be obtained as
\begin{eqnarray}\label{eq:down_approx_oma}
&&\hspace{-0.7cm}\bar{E}_{\oo,k,n}^\cc(\pmb{z}_n;\pmb{z}_n(v)) \triangleq E_{\oo,k,n}^\cc(L_{k,n}^\cc, \pmb{p}^\cc_n; L_{k,n}^\cc(v), \pmb{p}^\cc_n(v))\nonumber\\
&&\hspace{-0.7cm} =\frac{N_0 B \Delta/K}{2 g_0} \left[\left(2^{\frac{L_{k,n}^\cc}{B\Delta/K}}-1 \right.\right.\nonumber\\
&&\hspace{-0.7cm} \left.\left. +\left(x^\cc_n-x^\m_k\right)^2+\left(y^\cc_n-y^\m_k\right)^2+H^2\right)^2-\left(2^{\frac{L_{k,n}^\cc(v)}{B\Delta/K}}-1\right)^2\right.\nonumber\\
&& \hspace{-0.7cm}\left. - \left(\left(x^\cc_n(v)-x^\m_k\right)^2+\left(y^\cc_n(v)-y^\m_k\right)^2+H^2\right)^2\right]\nonumber\\
&&\hspace{-0.7cm} -\frac{N_0 \ln 2}{g_0}2^{\frac{L_{k,n}^\cc(v)}{B\Delta/K}} \left(2^{\frac{L_{k,n}^\cc(v)}{B\Delta/K}}-1\right)\left(L_{k,n}^\cc-L_{k,n}^\cc(v)\right)\nonumber\\
&&\hspace{-0.7cm} -\frac{2N_0 B \Delta/K}{g_0}\left(\left(x^\cc_n(v)-x^\m_k\right)^2+\left(y^\cc_n(v)-y^\m_k\right)^2+H^2\right)\nonumber\\
&& \hspace{-0.7cm}\left(\left(x^\cc_n(v)-x^\m_k\right)\left(x^\cc_n-x^\cc_n(v)\right) + \left(y^\cc_n(v)-y^\m_k\right)\left(y^\cc_n-y^\cc_n(v)\right) \right).\nonumber\\
\end{eqnarray}  

\section{Derivations of (\ref{eq:noma_approx})}\label{app:noma}
Here, for a given $\pmb{z}(v) \in \mathcal{X}$ with the feasible set $\mathcal{X}$ of problem (\ref{eq:opt_slack}), we derive the convex upper bounds of $h_{k,n}^\m(\pmb{z}_n;\pmb{z}_n(v))$ and $\hat{E}_{\nn,k,n}^\cc(\pmb{z}_n;\pmb{z}_n(v))$ in (\ref{eq:noma_approx}) similarly with Appendix \ref{app:oma} based on Lemma 2.

We can rewrite the non-convex function $h_{k,n}^\m(L_{k,n}^\m, \alpha_{-k,n})$ of (\ref{eq:alpha}) as 
\begin{eqnarray}
&&\hspace{-1.2cm}h_{k,n}^\m(\pmb{z}_n) \triangleq h_{k,n}^\m(L_{k,n}^\m, \alpha_{-k,n})=N_0 B\Delta\left(2^{\frac{L_{k,n}^\m}{B\Delta}}-1\right) \nonumber\\
&&\hspace{0cm} + \frac{1}{2}\left(2^{\frac{L_{k,n}^\m}{B\Delta}}-1+\sum_{k'=1, k'\neq k}^{K} \alpha_{k',n}\right)^2\nonumber\\
&&\hspace{0cm}-\frac{1}{2}\left(\left(2^{\frac{L_{k,n}^\m}{B\Delta}}-1\right)^2+\left(\sum_{k'=1, k'\neq k}^{K} \alpha_{k',n}\right)^2\right),
\end{eqnarray}
whose convex upper bound is given as 
\begin{eqnarray}\label{eq:h_noma_approx}
&&\hspace{-0.9cm} \bar{h}_{k,n}^\m(\pmb{z}_n; \pmb{z}_n(v)) \triangleq \bar{h}_{k,n}^\m(L_{k,n}^\m, \alpha_{-k,n};L_{k,n}^\m(v), \alpha_{-k,n}(v))\nonumber\\
&&\hspace{-0.9cm}=N_0 B\Delta\left(2^{\frac{L_{k,n}^\m}{B\Delta}}-1\right) \nonumber\\
&&\hspace{-0.9cm} + \frac{1}{2}\left[\left(2^{\frac{L_{k,n}^\m}{B\Delta}}-1+\sum_{k'=1, k'\neq k}^{K} \alpha_{k',n}\right)^2 - \left(2^{\frac{L_{k,n}^\m(v)}{B\Delta}}-1\right)^2 \right.\nonumber\\
&&\hspace{-0.9cm} -\left.\left(\sum_{k'=1, k'\neq k}^{K} \alpha_{k',n}(v)\right)^2\right]\nonumber\\
&&\hspace{-0.9cm} -\frac{\ln 2}{B\Delta}2^{\frac{L_{k,n}^\m(v)}{B\Delta}}\left(2^{\frac{L_{k,n}^\m(v)}{B\Delta}}-1\right)\left(L_{k,n}^\m-L_{k,n}^\m(v)\right)\nonumber\\
&&\hspace{-0.9cm}- \left(\sum_{k'=1, k'\neq k}^{K} \alpha_{k',n}(v)\right)\left(\sum_{k'=1, k'\neq k}^{K}\alpha_{k',n}-\alpha_{k',n}(v)\right).
\end{eqnarray}

Similarly, the non-convex function $\hat{E}_{\nn,k,n}^\cc(L_{k,n}^\cc, \pmb{p}^\cc_n, \beta_{-k,n})$ in the constraint (\ref{eq:beta}) can be expressed as
\begin{eqnarray}\label{eq:down_DC}
&&\hspace{-0.8cm} \hat{E}_{\nn,k,n}^\cc(\pmb{z}_n) \triangleq \hat{E}_{\nn,k,n}^\cc(L_{k,n}^\cc, \pmb{p}^\cc_n, \beta_{-k,n})\nonumber\\
&&\hspace{-0.8cm}= \frac{1}{2}\left[\frac{N_0 B\Delta}{g_0}\left(2^{\frac{L_{k,n}^\cc}{B\Delta}}-1+\left(x^\cc_n-x^\m_k\right)^2 \right.\right.\nonumber\\
&&\hspace{-0.8cm} \left.\left. +\left(y^\cc_n-y^\m_k\right)^2+H^2\right)^2 + \left(2^{\frac{L_{k,n}^\cc}{B\Delta}}-1+\sum_{k'=1, k'\neq k}^{K} \beta_{k',n}\right)^2\right]\nonumber\\
&&\hspace{-0.8cm}-\frac{1}{2}\left[\left(\frac{N_0 B\Delta}{g_0}+1\right)\left(2^{\frac{L_{k,n}^\cc}{B\Delta}}-1\right)^2 + \left(\sum_{k'=1, k'\neq k}^{K} \beta_{k',n}\right)^2\right.\nonumber\\
&&\hspace{-0.8cm}\left. + \frac{N_0 B\Delta}{g_0}\left(\left(x^\cc_n-x^\m_k\right)^2+\left(y^\cc_n-y^\m_k\right)^2+H^2\right)^2\right],
\end{eqnarray}
which is upper bounded by the convex surrogate function to linearize the concave parts of $\hat{E}_{\nn,k,n}^\cc(\pmb{z}_n)$ as
\begin{eqnarray*}\label{eq:up_noma_approx}
&&\hspace{-0.7cm} \bar{E}_{\nn,k,n}^\cc(\pmb{z}_n; \pmb{z}_n(v)) \nonumber\\
&&\hspace{-0.7cm}\triangleq \bar{E}_{\nn,k,n}^\cc(L_{k,n}^\cc, \pmb{p}^\cc_n, \beta_{-k,n};L_{k,n}^\cc(v), \pmb{p}^\cc_n(v), \beta_{-k,n}(v))\nonumber\\
&&\hspace{-0.7cm}= \frac{1}{2}\left[\frac{N_0 B\Delta}{g_0}\left(2^{\frac{L_{k,n}^\cc}{B\Delta}}-1+\left(x^\cc_n-x^\m_k\right)^2 \right.\right.\nonumber\\
&&\hspace{-0.7cm}\left.\left. + \left(y^\cc_n-y^\m_k\right)^2+H^2\right)^2 + \left(2^{\frac{L_{k,n}^\cc}{B\Delta}}-1+\sum_{k'=1, k'\neq k}^{K} \beta_{k',n}\right)^2 \right.\nonumber\\
&&\hspace{-0.7cm}\left. - \left(\frac{N_0 B\Delta}{g_0}+1\right)\left(2^{\frac{L_{k,n}^\cc(v)}{B\Delta}}-1\right)^2 \right.\nonumber\\
&&\hspace{-0.7cm}\left.  - \left(\sum_{k'=1, k'\neq k}^{K} \beta_{k',n}(v)\right)^2 - \frac{N_0 B\Delta}{g_0}\left(\left(x^\cc_n(v)-x^\m_k\right)^2 \right.\right.\nonumber\\
&&\hspace{-0.7cm}\left.\left. +\left(y^\cc_n(v)-y^\m_k\right)^2+H^2\right)^2\right]
\end{eqnarray*}
\begin{eqnarray}\label{eq:up_noma_approx}
&&\hspace{-2.4cm}- \ln 2\left(\frac{N_0}{g_0} + \frac{1}{B\Delta} \right)2^{\frac{L_{k,n}^\cc(v)}{B\Delta}}\left(2^{\frac{L_{k,n}^\cc(v)}{B\Delta}}-1\right)\nonumber\\
&&\hspace{-2.4cm} \times \left(L_{k,n}^\cc-L_{k,n}^\cc(v)\right) -\left(\sum_{k'=1, k'\neq k}^{K} \beta_{k',n}(v)\right)\nonumber\\
&&\hspace{-2.4cm} \times \left(\sum_{k'=1, k'\neq k}^{K}\beta_{k',n}-\beta_{k',n}(v)\right)-\frac{2N_0 B \Delta}{g_0} \nonumber\\
&&\hspace{-2.4cm} \times \left(\left(x^\cc_n(v)-x^\m_k\right)^2 +\left(y^\cc_n(v)-y^\m_k\right)^2+H^2\right) \nonumber\\
&&\hspace{-2.4cm} \times \left(\left(x^\cc_n(v)-x^\m_k\right)\left(x^\cc_n-x^\cc_n(v)\right) \right.\nonumber\\
&&\hspace{-2.4cm} \left.+ \left(y^\cc_n(v)-y^\m_k\right)\left(y^\cc_n-y^\cc_n(v)\right) \right).
\end{eqnarray}

\section{Derivations of Model 2 in (\ref{eq:pro})}\label{app:Ef}
Here, following \cite{Zeng16Arxiv2, Filippone06, Leishman06, Kong09AIAA}, we briefly discuss the propulsive energy consumption model (\ref{eq:pro}) which can be applied for both fixed-wing and rotary-wing UAV of weight $W = Mg$. For a fixed-wing UAV with initial and final velocity constraint (\ref{eq:initv}), the propulsion energy consumption is upper bounded by (\ref{eq:pro}), where $\kappa_1=0.5\rho C_{D_0} S_r\Delta$ and $\kappa_2=2M^2g^2\Delta /(\pi e_0 A_R \rho S_r)$ are derived by following \cite[Eq. (56)]{Zeng16Arxiv2}; $\rho$ is the air density in kg/m$^3$; $C_{D_0}$ is the zero-lift drag coefficient; $S_r$ is a reference area; $e_0$ is the Oswald efficiency; and $A_R$ is the aspect ratio of the wing. For a rotary-wing UAV, the power $P_F$ required for constant-height flight with speed $\|\pmb{v}^{\text{c}}\|$ can be approximated as \cite{Filippone06, Leishman06, Kong09AIAA}
\begin{equation}\label{eq:pf}
P_F \approx P_0 + P_p + P_i,
\end{equation}
where $P_0$ is the so called profile power, which is the power spent to turn the rotors and overcome the rotor aerodynamic drag force; $P_p$ is the so called parasitic power, which is the power required to overcome parasite drag; and $P_i$ is the so called induced power, which is the power required to produce lift by moving a mass of air through the disk at the induced velocity. In (\ref{eq:pf}), although the profile power $P_0$ is a function of flight speed $\|\pmb{v}^{\text{c}}\|$, its contribution is constant in low-speed flight and small compared to the other components, and is hence generally neglected. Moreover, following references \cite{Filippone06, Leishman06, Kong09AIAA}, the other two components in (\ref{eq:pf}) can be modeled as     
\begin{subequations}\label{eq:pf2}
\begin{eqnarray}
P_F(\pmb{v}^{\text{c}}, \pmb{a}^{\text{c}}) &\approx& 0.5\rho C_{D_f} S_r \|\pmb{v}^{\text{c}}\|^3 + \frac{\epsilon \|\pmb{T}\|^2}{2\rho A \|\pmb{v}^{\text{c}}\|} \\
&=& \frac{\kappa_1}{\Delta}\|\pmb{v}^{\text{c}}\|^3 + \frac{\kappa_2}{\Delta\|\pmb{v}^{\text{c}}\|}\left(1+\frac{\|\pmb{a}^{\text{c}}\|^2}{g^2}\right),
\end{eqnarray}
\end{subequations}
where $\kappa_1=0.5\rho C_{D_f} S_r\Delta$ and $\kappa_2=\epsilon M^2g^2\Delta /(2 \rho A)$; $\pmb{a}^{\text{c}}$ is the UAV's acceleration vector; $C_{D_f}$ is the drag coefficient based on the reference area $S_r$; $A$ is the area of the main rotor disk; $\epsilon$ is the induced power factor; and $\pmb{T}$ is the total required thrust, which can be calculated as $\|\pmb{T}\|^2 = W^2 (1 + \|\pmb{a}^{\text{c}}\|^2/g^2)$ for constant-height flight. For a trajectory $\pmb{p}^{\text{c}}(t)$, velocity $\pmb{v}^{\text{c}}(t)$ and acceleration $\pmb{a}^{\text{c}}(t)$, the total propulsion energy is then given by integrating (\ref{eq:pf2}) over time 
\begin{equation}\label{eq:Ef_con}
E^{\text{c}}_F(\pmb{v}^{\text{c}}(t), \pmb{a}^{\text{c}}(t)) = \int P_F(\pmb{v}^{\text{c}}(t), \pmb{a}^{\text{c}}(t)) dt.
\end{equation}
By applying the discrete linear state-space approximation in \cite{Zeng16Arxiv2} to (\ref{eq:Ef_con}), the rotary-wing UAV's propulsion energy consumption at the $n$th frame can be also derived as Model 2 in (\ref{eq:pro}).

\bibliographystyle{IEEEtran}
\bibliography{JSAref}

\end{document}